%% file: arxiv.tex
\documentclass[acmtog]{acmart}

\usepackage[ruled]{algorithm2e} % For algorithms

\SetAlFnt{\small}
\SetAlCapFnt{\small}
\SetAlCapNameFnt{\small}
\SetAlCapHSkip{0pt}

\usepackage[capitalise]{cleveref}
\usepackage{listings}
\crefname{lstlisting}{Listing}{Listings}
\Crefname{lstlisting}{Listing}{Listings}
\lstdefinestyle{promptbox}{
  basicstyle=\ttfamily\footnotesize,
  breaklines=true,
  breakatwhitespace=false,
  columns=fullflexible,
  captionpos=t,
  frame=single,
  framerule=0.3pt,
  backgroundcolor=\color{black!3},
  keepspaces=true,
  showstringspaces=false,
  upquote=true,
  xleftmargin=0.5em,
  xrightmargin=0.5em,
  aboveskip=0.75\baselineskip,
  belowskip=0.75\baselineskip
}

\acmJournal{TOG}

\copyrightyear{2026}
\acmYear{2026}
\setcopyright{cc}
\setcctype{by}
\acmConference[SA Conference Papers '26]{SIGGRAPH Asia 2026 Conference Papers}{December 01--04, 2026}{Kuala Lumpur, Malaysia}
\acmBooktitle{SIGGRAPH Asia 2026 Conference Papers (SA Conference Papers '26), December 01--04, 2026, Kuala Lumpur, Malaysia}
\acmDOI{10.1145/3829340.3842289}
\acmISBN{979-8-4007-2842-6/2026/12}

\begin{document}
\title[GALA: Geometry-Aware Language Model for Controllable Object Arrangement]{GALA: \texorpdfstring{\underline{G}eometry-\underline{A}ware \underline{L}anguage Model for Controllable Object \underline{A}rrangement}{Geometry-Aware Language Model for Controllable Object Arrangement}}

\author{Chucheng Xiang}
\orcid{https://orcid.org/0009-0000-3310-5528}
\affiliation{%
  \institution{University of Science and Technology of China}
  \city{Hefei}
  \country{China}}
\affiliation{%
  \institution{Tencent}
  \city{Shenzhen}
  \country{China}}
\email{xcc2020@mail.ustc.edu.cn}

\author{Runze Wang}
\orcid{https://orcid.org/0009-0002-0507-8715}
\affiliation{%
  \institution{University of Science and Technology of China}
  \city{Hefei}
  \country{China}}
\affiliation{%
  \institution{Tencent}
  \city{Shenzhen}
  \country{China}}
\email{runzewang@mail.ustc.edu.cn}

\author{Zhi Deng}
\authornote{Corresponding authors.}
\orcid{https://orcid.org/0000-0003-4582-4874}
\affiliation{%
  \institution{Tencent}
  \city{Shenzhen}
  \country{China}}
\email{abeldeng@tencent.com}

\author{Ruchao Bao}
\orcid{https://orcid.org/0009-0002-4536-1953}
\affiliation{%
  \institution{University of Science and Technology of China}
  \city{Hefei}
  \country{China}}
\affiliation{%
  \institution{Tencent}
  \city{Shenzhen}
  \country{China}}
\email{iambrc@mail.ustc.edu.cn}

\author{Yuanwei Zhang}
\orcid{https://orcid.org/0009-0000-8388-5665}
\affiliation{%
  \institution{Tsinghua University}
  \city{Shenzhen}
  \country{China}}
\affiliation{%
  \institution{Tencent}
  \city{Shenzhen}
  \country{China}}
\email{yuanwei-24@mails.tsinghua.edu.cn}

\author{Yuxuan Xie}
\orcid{https://orcid.org/0000-0002-9877-4594}
\affiliation{%
  \institution{Tencent}
  \city{Shenzhen}
  \country{China}}
\email{yuxuan.xie@hotmail.com}

\author{Hanliu Wang}
\orcid{https://orcid.org/0009-0001-4781-5077}
\affiliation{%
  \institution{Tencent}
  \city{Shenzhen}
  \country{China}}
\email{wanghanliu97@gmail.com}

\author{Liangzhen Fei}
\orcid{https://orcid.org/0009-0006-0919-9273}
\affiliation{%
  \institution{University of Science and Technology of China}
  \city{Hefei}
  \country{China}}
\affiliation{%
  \institution{Tencent}
  \city{Shenzhen}
  \country{China}}
\email{flzcn6@mail.ustc.edu.cn}

\author{Wenzheng Wu}
\orcid{https://orcid.org/0009-0005-4192-8176}
\affiliation{%
  \institution{University of Science and Technology of China}
  \city{Hefei}
  \country{China}}
\affiliation{%
  \institution{Tencent}
  \city{Shenzhen}
  \country{China}}
\email{wuwzh@mail.ustc.edu.cn}

\author{Cheng Wan}
\orcid{https://orcid.org/0009-0000-6350-4925}
\affiliation{%
  \institution{Hong Kong University of Science and Technology}
  \city{Hong Kong}
  \country{China}}
\affiliation{%
  \institution{Tencent}
  \city{Shenzhen}
  \country{China}}
\email{wanc7147@gmail.com}

\author{Peifeng Li}
\orcid{https://orcid.org/0009-0002-9847-6017}
\affiliation{%
  \institution{University of Science and Technology of China}
  \city{Hefei}
  \country{China}}
\affiliation{%
  \institution{Tencent}
  \city{Shenzhen}
  \country{China}}
\email{lipf6@mail.ustc.edu.cn}

\author{Zhongyuan Liu}
\orcid{https://orcid.org/0000-0003-1601-0038}
\affiliation{%
  \institution{Tencent}
  \city{Shenzhen}
  \country{China}}
\email{lockliu@tencent.com}

\author{Ligang Liu}
\authornotemark[1]
\orcid{https://orcid.org/0000-0003-4352-1431}
\affiliation{%
  \institution{University of Science and Technology of China}
  \city{Hefei}
  \country{China}}
\email{lgliu@ustc.edu.cn}

\renewcommand\shortauthors{Xiang et al.}

\begin{abstract}
Reasoning about fine-grained placement constraints, such as support and containment, is essential for realistic 3D scene creation.
It remains a significant challenge for existing methods that rely on simplified object representations like oriented bounding box (OBB), which are blind to the detailed geometry.
To address this limitation, we introduce GALA, a geometry-aware language model with explicit point cloud perception for controllable object placement.
Our model performs autoregressively by predicting the 6D pose of one target object at a time conditioned on the scene context and placement instruction.
We utilize a two-stage training strategy: the first stage aligns geometric and textual modalities through pretraining tasks, and the second stage fine-tunes the model for placement.
To facilitate this task, we also curate a multi-source scene dataset covering diverse room types and fine-grained furniture arrangements, which will be open-sourced for community benefit.
Experiments on our benchmark show our model significantly outperforms baselines without explicit geometric reasoning.
We further develop a multi-stage plan--place--verify agent pipeline that uses the model for text-driven scene generation.
\par\noindent\textbf{Project Page:} \url{https://xccelephant.github.io/paper/gala/}
\end{abstract}




\input{sections/figs/teaser}

\maketitle

\input{sections/introduction}
\input{sections/related_work}
\input{sections/method}
\input{sections/experiments}
\input{sections/conclusions}

\begin{acks}
This work was supported by the
\grantsponsor{NSFC}{National Natural Science Foundation of China}{https://www.nsfc.gov.cn/}
under Grant \grantnum{NSFC}{62025207}.
We thank the reviewers for their time and constructive feedback.
\end{acks}

\clearpage
\bibliographystyle{ACM-Reference-Format}
\bibliography{refer}

\clearpage
\appendix
\twocolumn[
  \begin{center}
    {\LARGE\sffamily\bfseries Supplementary Material\par}
    \bigskip
  \end{center}
]
\input{sections/supplement_body}

\end{document}

%% file: sections/figs/teaser.tex
\begin{teaserfigure}
  \centering
  \includegraphics[width=0.99\textwidth]{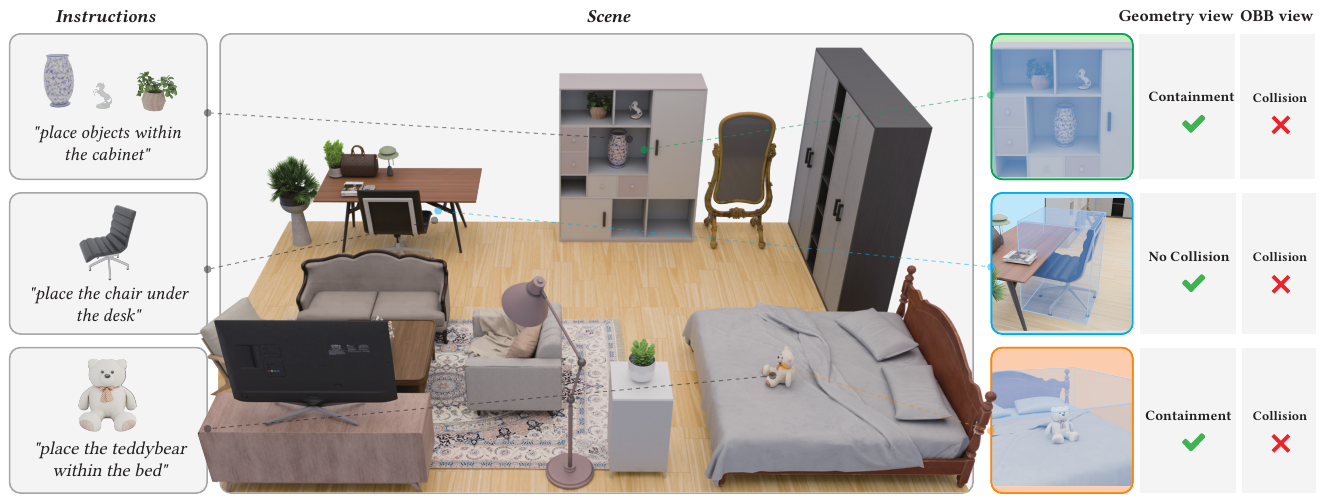}
  \caption{Left: input objects and fine-grained placement instructions. Middle: plausible layouts that follow the instructions. Right: zoom-in views show how GALA goes beyond oriented bounding boxes by perceiving geometry to produce correct support, containment, and collision-free placements.}
  \Description{Teaser figure showing input placement goals, instruction-conditioned layout results from a bounding-box baseline and GALA, and zoom-in views of GALA placements with correct support, containment, and collision-free geometry.}
  \label{fig:teaser}
\end{teaserfigure}

%% file: sections/introduction.tex
\section{Introduction}
\label{sec:intro}

Generating realistic 3D scenes is important in game development, film production, architectural design, and embodied AI training, where virtual environments must be both physically plausible and semantically meaningful.
Scene construction can be viewed as an iterative process of single-object placement: each step inserts a new object conditioned on the scene built so far. This single-object placement problem is the focus of this work.
Placing a new object requires understanding both high-level semantic constraints, such as user instructions and object relationships, and low-level geometric constraints, such as collision avoidance, support, and containment.

However, most existing methods place objects only at the level of oriented bounding box (OBB): each object is represented by its center, size, and rotation. 
This abstraction is sufficient for coarse arrangements, but it struggles with the geometric constraints in realistic scenes, as shown in \cref{fig:teaser}.
Three basic failure modes are common in practice. First, a chair placed under a desk must clear the desktop with its seat while avoiding the table legs.
Second, small objects like books are often placed inside shelves whose internal cavities are invisible to the OBB.
Third, cushions placed on curved surfaces, such as sofas, require reasoning about support heights that vary across the contact region.
In all three cases, OBB-only methods cannot encode the fine-grained constraints, leading to either collisions (in OBB space), floating, or misalignments with the underlying geometry.

We view this limitation as primarily a geometry-awareness problem rather than an algorithmic one.
A natural solution is to bring geometric information into the placement problem through a 3D representation of the objects and the scene.
Among possible 3D representations, point clouds offer a favorable trade-off: they are lightweight, easy to obtain, and uniform across asset libraries, and recent work on 3D multimodal LLMs~\cite{pointllm,3dllm,mao2026spatiallm} shows that LLMs can consume point clouds directly.
Our core hypothesis is simple: LLM-based placement models augmented with point clouds can handle fine-grained arrangements beyond OBB reasoning.

Building on this hypothesis, we introduce GALA, a geometry-aware placement model.
The model is a multimodal LLM built on a strong LLM backbone, extended with a point cloud encoder and a projector that maps geometry features into the LLM embedding space.
Given an existing scene, an object to be placed, and a placement instruction, the model predicts the 6D pose of the target object.
The instruction specifies the intended relation and makes object insertion controllable.
To make the point cloud tokens readable to the LLM, we adopt a two-stage training recipe: geometric pretraining aligns the projector, and task fine-tuning adapts the full model for object placement.
Existing scene datasets, such as 3D-FRONT~\cite{3dfront}, contain noisy scenes and provide limited examples of fine-grained placement.
We therefore curate a multi-source scene dataset from 3D-FRONT, Imaginarium~\cite{zhu2025imaginarium}, SpatialLM~\cite{mao2026spatiallm}, and an artist-crafted set of 300 scenes, followed by filtering noisy layouts and adding fine-grained placement examples.
Finally, to make our model easier to use and support automated scene generation, we develop a multi-stage agent pipeline that converts textual user prompts into complete 3D scenes through object planning, instruction generation, and geometry-aware placement.

We summarize our contributions as follows.

\begin{itemize}
  \item We introduce a geometry-aware language model for controllable object pose prediction conditioned on placement instruction and point cloud geometry, enabling fine-grained placement beyond OBB-based reasoning.
  \item We curate a multi-source high-quality scene dataset, combining public scenes and artist-crafted scenes.
  The dataset is physically and visually verified, contains complex spatial relations, and will be open-sourced for community benefit.
  \item We develop a plan--place--verify agent system for automatic text-driven scene generation, integrating object planning, instruction generation, and geometry-aware placement.
\end{itemize}

%% file: sections/related_work.tex
\section{Related Work}
\label{sec:related}

\input{sections/figs/model_overview}

\subsection{Data-driven Scene Synthesis}
Early learning-based approaches represent an indoor scene as a set of oriented bounding boxes and learn the joint distribution of object classes and box parameters, including convolutional~\cite{wang2018deep, wang2019planit} or recursive~\cite{li2019grains} priors, autoregressive models~\cite{atiss,bucher2025respace,feng2026repurposing}, and diffusion-based models~\cite{diffuscene}.
Other lines of work model scenes as graphs and predict pairwise relations~\cite{scenegraphnet,gao2023scenehgn,zhai2023commonscenes,zhai2024echoscene,lin2024instructscene}.
These methods reason at the OBB level, so they cannot express the geometric detail needed to slide a chair under a table or place a book on a specific shelf inside a cabinet. They also tie generation to the scene dataset distribution, which limits generalization to unseen room types.

\subsection{LLM- and VLM-based Scene Layout}
A recent line of work treats LLMs and VLMs as priors over indoor scenes.
LayoutGPT~\cite{layoutgpt} prompts an LLM with in-context examples to output object positions and sizes from a scene description, while LayoutVLM~\cite{layoutvlm} leverages the visual abilities of modern VLMs to generate scenes from text.
Holodeck~\cite{holodeck} decomposes scene generation into LLM-driven planning, asset retrieval, and constraint-based placement.
Because these model outputs are still coarse layouts, several methods add explicit optimization or solvers after the language-model stage.
\cite{gumin2025procedural, wu2026scene} generate indoor scenes through LLM-generated programs followed by layout optimization.
Co-Layout~\cite{xiang2026co} couples an LLM-driven agent workflow with grid-based integer programming to jointly optimize room layouts and furniture placements.
FirePlace~\cite{huang2025fireplace} similarly uses VLM-generated constraints and a solver for object placement.
Closer to our setting, PlaceIt3D~\cite{placeit3d} trains a 3D LLM for language-guided placement on reconstructed scanned scenes, but predicts a placement mask rather than a numeric 6D pose, and does not consider fine-grained containment relations.
These methods show that pretrained foundation models carry useful priors over spatial common sense, but they still make geometric decisions with coarse boxes, constraints, or solvers.
As a result, fine-grained placements that depend on object shape are not handled through native 3D geometric perception.

\subsection{3D Multimodal Large Language Models}
Recent 3D multimodal LLMs extend language models from image-text reasoning to point-cloud and scene-level understanding.
Early systems such as 3D-LLM~\cite{3dllm} and PointLLM~\cite{pointllm} show that LLMs can describe 3D content, answer questions, and follow instructions when supplied with 3D inputs.
This direction has since expanded to more structured 3D tasks: 3D-LLaVA~\cite{deng20253d} studies general 3D instruction following, SegPoint~\cite{segpoint} and Reason3D~\cite{reason3d} use LLMs for point-level segmentation from referring or reasoning queries, and SpatialLM~\cite{mao2026spatiallm} targets structured indoor scene understanding.
Recent 3D VLM work further studies spatial reasoning itself, such as SpatialStack~\cite{spatialstack}, which focuses on point-language reasoning for 3D spatial relations.
These works make 3D geometry accessible to foundation models, but their outputs are mostly language responses, masks, or structured scene descriptions rather than precise poses for inserting a new object into an existing scene.
Our work uses the 3D multimodal formulation for object placement, where the model must map scene geometry, target object geometry, and a placement instruction to a numeric 6D pose.

%% file: sections/figs/model_overview.tex
\begin{figure*}[t]
    \centering
    \includegraphics[width=0.95\textwidth]{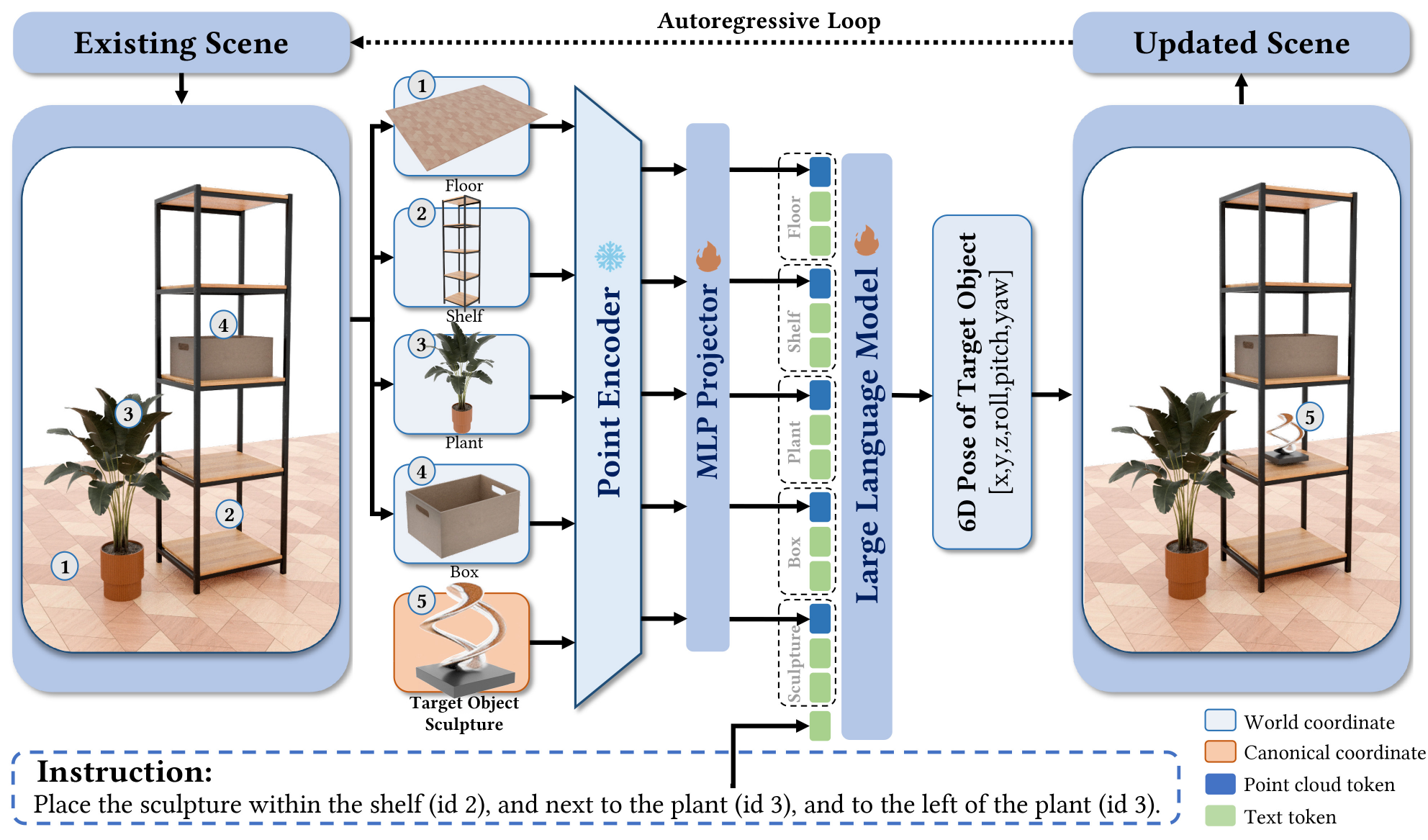}
    \caption{Overview of the geometry-aware placement model.
    Given the scene boundary, the already-placed objects (with poses, sizes, semantic categories, and point clouds), the new object (with size, semantic category, and point cloud), and a placement instruction, the multimodal LLM outputs a numeric 6D pose.
    The model performs placement autoregressively, adding each predicted object to the scene context before predicting the next one.}
    \Description{
      Overview of the geometry-aware placement model.
    }
    \label{fig:model}
\end{figure*}

%% file: sections/method.tex
\input{sections/figs/dataset_pipeline}
\section{Method}
\label{sec:method}
\paragraph{Problem Formulation} We study a single-object placement task in a bounded scene region.
The scene region is described by a 2D footprint polygon $\mathcal{B} \subset \mathbb{R}^2$, together with a floor point cloud $P_{\mathcal{B}}$ sampled on the area it encloses.
A scene consists of a set of existing object instances $\mathcal{O}_{\mathrm{exist}} = \{O_i\}_{i=0}^{N-1}$.
Each existing object $O_i$ is associated with an identifier $i$, a semantic category string $c_i$, an axis-aligned size $\boldsymbol{s}_i=(w_i, d_i, h_i)$, and a point cloud $P_i$.
Once placed, it has a 6D pose $\boldsymbol{p}_i = (x_i, y_i, z_i, \phi_i, \theta_i, \psi_i)$, where $(x_i, y_i, z_i)$ denotes the translation, and $(\phi_i, \theta_i, \psi_i)$ denotes the roll, pitch, and yaw angles, respectively.
Here we use a world frame with $+X$ pointing right, $+Y$ pointing forward, and $+Z$ pointing up. By convention, an object faces the $-Y$ direction when $\psi_i = 0$, and positive yaw values are defined counterclockwise.
The object to be placed, $O_{\mathrm{tar}}$, is specified by its category $c_{\mathrm{tar}}$, size $\boldsymbol{s}_{\mathrm{tar}}$, and point cloud $P_{\mathrm{tar}}$, while the placement instruction $\boldsymbol{I}$ constrains the desired spatial relations to the scene and reference objects, for example \emph{``place the table against the west wall, in front of the sofa (id 0), and to the left of the wardrobe (id 3)''}.
We want to train a predictor $f_\theta$ to output the 6D pose of the target object,
\begin{equation}
  \boldsymbol{p}_{\mathrm{tar}} \;=\; f_\theta\!\left(\,\mathcal{B},\; P_{\mathcal{B}},\;
  \mathcal{O}_{\mathrm{exist}},\; c_{\mathrm{tar}},\; \boldsymbol{s}_{\mathrm{tar}},\; P_{\mathrm{tar}},\; \boldsymbol{I}\,\right).
  \label{eq:problem}
\end{equation}

\subsection{Geometry-Aware Multimodal Architecture}
\label{sec:method:model}

As illustrated in \cref{fig:model}, we instantiate $f_\theta$ as a multimodal LLM whose text backbone is a pretrained causal LLM and whose geometric branch is a 3D point cloud encoder $\mathcal{E}$ followed by a projector $g$.

\paragraph{Point cloud perception module} To achieve precise geometric perception of 3D scenes, we employ Utonia~\cite{zhang2026utonia}, a state-of-the-art, self-supervised point transformer model. Its pretraining on diverse 3D data is crucial, as it mitigates the need for domain-specific priors, allowing the model to capture scene-level scale and fine-grained object geometry.
The encoder $\mathcal{E}$ takes a point cloud $P \in \mathbb{R}^{N_p\times C}$ with $N_p$ input points and $C$ channels (coordinates, color, and normals) as input, and returns a feature matrix $\mathcal{E}(P)\in\mathbb{R}^{M\times d_e}$, where $M$ is the number of output point features, typically with $M < N_p$, and $d_e$ is the feature dimension.

To avoid feature conflation from a single scene-level point cloud, we encode each object's geometry independently. This gives the model separate point features for each instance.

\paragraph{Geometric-to-text projection} We use a projector to align geometry features with the text embedding space.
The projector $g: \mathbb{R}^{d_e}\rightarrow \mathbb{R}^{d_h}$ is parameterized by a two-layer MLP with GELU activation, where $d_h$ is the LLM hidden dimension.
For the point cloud $P_i$ of $i$-th object $O_{i}$, the coupled encoder-projector architecture generates a block of geometric token embeddings,
\begin{equation}
  \mathbf{H}^{(i)} = g(\mathcal{E}(P_i))
  = \left(\boldsymbol{h}^{(i)}_1,\dots,\boldsymbol{h}^{(i)}_{M_i}\right),
  \quad \boldsymbol{h}^{(i)}_m \in \mathbb{R}^{d_h}.
  \label{eq:point-token-block}
\end{equation}

\paragraph{Interleaved multimodal context} We interleave the projected point-cloud tokens with the text tokens according to their positions in the prompt.
The prompt verbalizes the inputs of \cref{eq:problem} as a sequence of text segments, each followed by the point-cloud token block of the geometry it introduces.
The leading segment $\mathbf{u}_{\mathcal{B}}$ describes the footprint polygon $\mathcal{B}$ and is followed by the floor block $\mathbf{H}^{(\mathcal{B})}=g(\mathcal{E}(P_{\mathcal{B}}))$.
Each existing object $O_i$ then contributes a segment $\mathbf{u}_i$ that lists its identifier, category, size, and pose $(i, c_i, \boldsymbol{s}_i, \boldsymbol{p}_i)$, followed by its geometry block $\mathbf{H}^{(i)}$, for $i=0,\dots,N-1$.
Segment $\mathbf{u}_{\mathrm{tar}}$ lists the target category and size $(c_{\mathrm{tar}}, \boldsymbol{s}_{\mathrm{tar}})$ but no pose, and is followed by $\mathbf{H}^{(\mathrm{tar})}=g(\mathcal{E}(P_{\mathrm{tar}}))$, while the final segment $\mathbf{u}_{\mathrm{ins}}$ carries the instruction $\boldsymbol{I}$.
The floor and existing-object point clouds are encoded in world coordinates, the target point cloud in its canonical coordinates.
For example, $\mathbf{u}_0$ reads \emph{``object 0: sofa; size [2.10, 0.90, 0.85]; pose [1.20, 2.00, 0.43, 0, 0, 90]; point cloud:} \texttt{<pc>}\emph{''}, where the placeholder \texttt{<pc>} is replaced by the projected tokens $\mathbf{H}^{(0)}$.
Writing $\mathbf{U}_j=\mathrm{TextEmb}(\mathbf{u}_j)$ for each segment embedding, we obtain the interleaved context sequence,
\begin{equation}
  \mathbf{X}_{\mathrm{ctx}}
  =
  \left[
  \mathbf{U}_{\mathcal{B}};
  \mathbf{H}^{(\mathcal{B})};
  \mathbf{U}_0;
  \mathbf{H}^{(0)};
  \cdots;
  \mathbf{U}_{N-1};
  \mathbf{H}^{(N-1)};
  \mathbf{U}_{\mathrm{tar}};
  \mathbf{H}^{(\mathrm{tar})};
  \mathbf{U}_{\mathrm{ins}}
  \right].
  \label{eq:context-embedding-sequence}
\end{equation}

The causal LLM prefills this sequence with standard self-attention, perceiving the scene context and the target geometry jointly before it autoregressively decodes the target pose.

\subsection{Dataset Construction}
\label{sec:method:dataset}

To train the geometry-aware placement model, we construct a high-quality scene dataset containing object semantics, point-level geometry, spatial relations, and 6D poses, following the pipeline in \cref{fig:dataset_pipeline}.
The dataset combines four complementary sources: two public datasets, 3D-FRONT~\cite{3dfront} and Imaginarium~\cite{zhu2025imaginarium}; a SpatialLM-PCG dataset generated from SpatialLM~\cite{mao2026spatiallm} scene annotations through procedural asset replacement and small-object enrichment; and an artist-crafted scene collection.
The PCG dataset increases the frequency and diversity of complex object relations, with construction details provided in the supplemental material.
As summarized in \cref{fig:dataset_distribution}, the dataset covers a wide range of scene complexities. SpatialLM-PCG and artist-crafted scenes provide many placement instructions and increase the number of support and containment relations.
\input{sections/figs/data_comparison}

\paragraph{Dataset filtering} As shown in \cref{fig:dataset_pipeline}, all sources are processed by a unified filtering pipeline before training.
Physics simulation and geometric validation correct or remove floating objects, severe interpenetrations, and out-of-bound placements.
We then render each candidate scene and use a VLM to score placement plausibility, scene coherence, visual quality, style consistency, and semantic reasonableness.
We discard samples failing these checks to reduce noisy supervision.

\paragraph{Scene data structuring} We then convert each retained scene into single-step autoregressive placement samples.
For each object, we extract its category, size, location, rotation, and point cloud, and derive spatial relations from oriented bounding boxes in the world coordinate frame.
We build a floor-rooted scene tree $\mathcal{T}$ from relations defining placement dependencies: \emph{abs} edges attach floor-supported objects to room-level location cues, while \emph{on}/\emph{within}/\emph{under} edges attach objects to their supporting, containing, or overhead parents.
These labels are geometric predicates rather than colloquial terms: \emph{within} requires the child OBB to lie inside the parent OBB, and \emph{on} requires the child bottom to contact the parent OBB top.
A breadth-first traversal maps this tree to the linear placement order $\boldsymbol{\pi}=\mathrm{BFS}(\mathcal{T})=(\pi_1,\dots,\pi_N)$.
For each target object in this order, the instruction is formed from its tree edge and augmented with nearby \emph{next to} relations to provide local distance, direction, and facing context, such as left/right and front/behind, relative to already placed objects.
The target 6D pose serves as the supervision signal.

\subsection{Two-Stage Training Strategies}
\label{sec:method:training}
Multimodal LLMs commonly separate modality alignment from downstream task fine-tuning: a lightweight projector is first trained to map visual or geometric features into the LLM token space, and the aligned model is then adapted to task-specific supervision~\cite{liu2023visual,pointllm,deng20253d}.
We follow this recipe because our point encoder, projector, and LLM operate in different feature spaces.
We first align the geometric and language modalities through auxiliary alignment tasks, and then fine-tune the model on autoregressive placement.

\paragraph{Training objective} Both stages use the standard causal language modeling objective.
For a training sample with context $\mathbf{X}_{\mathrm{ctx}}$, let $\mathbf{t}=(t_1,\dots,t_L)$ denote the complete task-specific target token sequence.
With teacher forcing, we optimize the negative log-likelihood of the target tokens,
\begin{equation}
  \mathcal{L}
  =
  - \sum_{\ell=1}^{L}
  \log p(t_\ell \mid \mathbf{X}_{\mathrm{ctx}}, t_{<\ell}).
  \label{eq:training-objective}
\end{equation}
\paragraph{Stage 1: geometric--language alignment} We first construct auxiliary alignment samples from the scene dataset.
Each sample pairs the floor point cloud or a set of object point clouds with a textual question answerable only from geometry.
The questions cover boundary descriptions, object captions, object centers, sizes, rotations, and pairwise distances or displacements.
Since Utonia stays frozen, this stage does not learn geometry from scratch: the geometric competence already comes from large-scale pretraining, and the projector only has to map the pretrained features into the language embedding space.
PAR3D~\cite{dai2026par3d} builds part-aware 3D reasoning on the same encoder, indicating that its features retain shape detail well beyond a bounding box.

\paragraph{Stage 2: autoregressive placement training} The aligned model is then fine-tuned on autoregressive placement samples.
Following the placement order $\boldsymbol{\pi}$, the model conditions at step $n$ on the ground-truth poses of $O_{\pi_1},\dots,O_{\pi_{n-1}}$ and uses the pose of $O_{\pi_n}$ as the supervision target.
We serialize each 6D pose $\boldsymbol{p}=(x,y,z,\phi,\theta,\psi)$ into a fixed-format textual sequence and tokenize it with the LLM tokenizer.
The point encoder is kept frozen, while the projector and LLM are jointly trained to combine the textual instruction with geometric context and predict a precise pose.
At inference time, placement proceeds autoregressively: at each step, the model conditions on previously predicted poses, generates a textual pose sequence, parses it into numeric pose parameters, updates the scene with the newly placed object, and advances to the next one.

\input{sections/figs/agent_pipeline}
\subsection{Agent Pipeline for Automated Scene Generation}
\label{sec:method:agent}
For text-driven scene generation, we design a multi-stage agent pipeline that wraps GALA in a plan-place-verify loop, as shown in \cref{fig:agent_pipeline}.
The pipeline separates semantic decisions from numeric ones: two off-the-shelf LLM agents decide what to place and how to describe it, while GALA, the only trained component, turns each description into a 6D pose.
The planner analyzes the user prompt once per scene and expands it into an ordered object list, giving each entry a category, a retrieval description, an estimated size, and the earlier objects its placement depends on.
Assets are retrieved from an asset library by description and size, and the list is ordered coarse to fine so that anchors precede the objects they support.
At each step, the instruction generator reads the current scene state, serialized as per-object identifiers, categories, sizes, and poses, together with the next object in the list.
It emits a placement instruction in the same relation vocabulary used for training, such as \emph{``place the mouse (id 14) on the mouse pad (id 13)''}, so that GALA receives the same kind of instruction at inference as during training.
GALA predicts the pose of the target object, after which a verifier checks the updated scene for out-of-boundary placements, collisions, floating objects, and unsatisfied relations, and sends any failure back to the instruction generator for the next iteration.
The loop terminates once the object list is exhausted; the planner and instruction generator prompts are listed in the supplemental material.
Gallery results from this agent pipeline are shown in \cref{fig:agent-gallery}, spanning living rooms, storage spaces, corridors, rooftops, and outdoor alleys.

%% file: sections/figs/dataset_pipeline.tex
\begin{figure*}[t]
  \centering
  \includegraphics[width=0.95\textwidth]{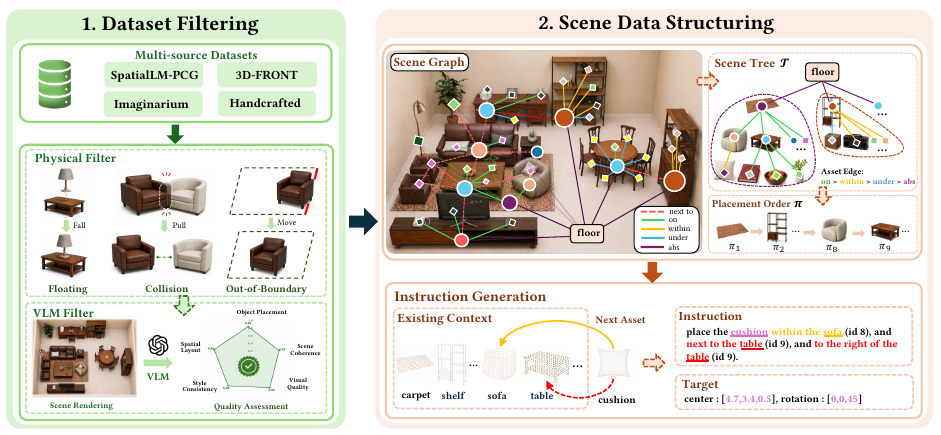}
  \caption{Dataset construction pipeline. Multi-source scenes are first cleaned by physical plausibility checks and VLM-based quality assessment. Each retained scene is parsed into a scene graph containing absolute room-location edges (\emph{abs}), dependency edges (\emph{on}, \emph{within}, \emph{under}), and proximity edges (\emph{next to}). We form a floor-rooted scene tree $\mathcal{T}$ by selecting dependency edges with priority \emph{on} $>$ \emph{within} $>$ \emph{under} $>$ \emph{abs}. We obtain the placement order $\boldsymbol{\pi}=\mathrm{BFS}(\mathcal{T})$ by traversing from the floor root, sorting nodes at the same depth by edge priority and footprint area. For each placement step, the tree edge and nearby proximity relations are verbalized into an instruction paired with per-object point clouds and the target 6D pose.}
  \Description{Pipeline for dataset filtering and scene data structuring,
  showing multi-source datasets, physical and VLM-based filtering, scene
  graph construction with absolute, dependency, and proximity relations,
  floor-rooted scene tree extraction with relation priority, BFS placement
  order generation, and instruction-target sample construction.}
  \label{fig:dataset_pipeline}
\end{figure*}

%% file: sections/figs/data_comparison.tex
\begin{figure}[t]
  \centering
  \includegraphics[width=\columnwidth]{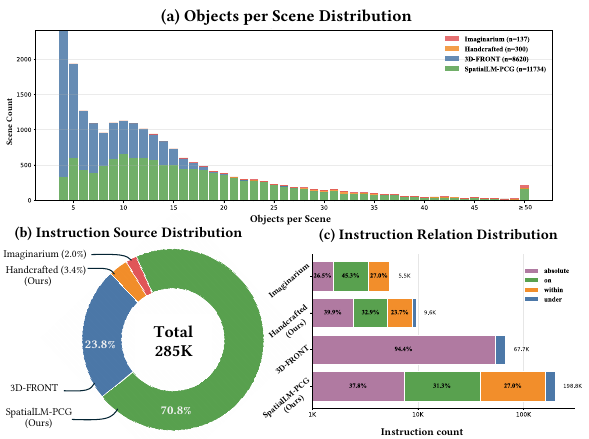}
  \caption{Dataset statistics. (a) Distribution of object counts per scene across data sources. (b) Proportion of placement instructions generated from each source. (c) Distribution of instruction relation types, including \emph{absolute}, \emph{on}, \emph{within}, and \emph{under}.}
  \Description{Dataset statistics showing objects per scene, instruction
  source proportions across Imaginarium, artist-crafted scenes, 3D-FRONT,
  and SpatialLM-PCG, and instruction relation distributions for absolute,
  on, within, and under relations.}
  \label{fig:dataset_distribution}
\end{figure}

%% file: sections/figs/agent_pipeline.tex
\begin{figure}[t]
  \centering
  \includegraphics[width=\linewidth]{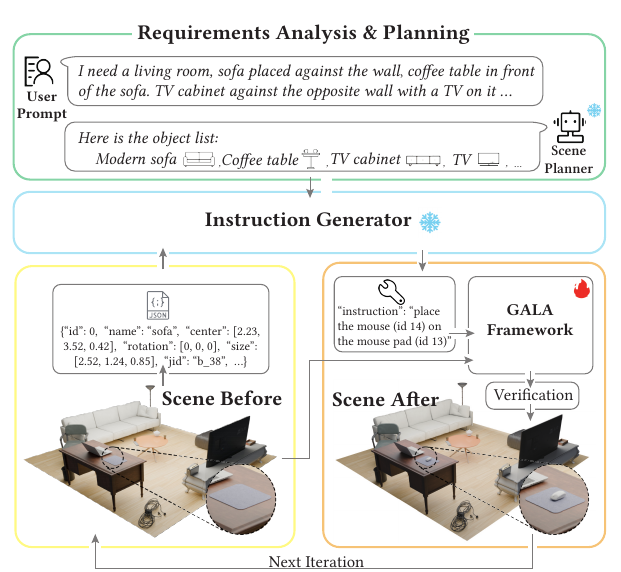}
  \caption{Agent pipeline for automated scene generation, where a planner turns a user prompt into object-level steps, GALA places each object, and a verifier feeds back failures for correction.}
  \Description{The agent pipeline converts a user prompt into object plans, generates placement instructions, uses GALA to predict poses, and verifies each updated scene before continuing.}
  \label{fig:agent_pipeline}
\end{figure}

%% file: sections/experiments.tex
\section{Experiments}
\label{sec:experiments}

\input{sections/figs/comparison_single_placement}

\subsection{Experimental Settings}
\label{sec:experiments:settings}

\paragraph{Datasets} We use the two-stage datasets produced by the construction pipeline in \cref{sec:method:dataset}.
To avoid leakage, we split scenes into train, validation, and test sets with stratified sampling over the source collections at an 18:1:1 ratio.
All evaluations are conducted on the test set.

\paragraph{Evaluation metrics}
We evaluate placements along two complementary axes: \emph{instruction following} and \emph{geometric sensitivity}.
Jointly, we deem a placement \textit{valid} if its pose is well-formed, satisfies all specified relations, and is collision-free and non-floating.
We report two success rates over this criterion: \textit{Instruction Valid Rate}, the overall validity across all test samples, and \textit{Within Valid Rate}, the validity on containment (\emph{within}) samples that serves as our primary geometry-sensitive metric since such relations demand local geometric reasoning.
We further report the \textit{Floating Rate} and the bottom-surface \textit{Support Ratio} to gauge physical support from the floor or existing objects.
Detailed metric definitions are provided in the supplemental material.

\paragraph{Implementation details} We implement our model using LLaMA-Factory~\cite{zheng2024llamafactory}, with Qwen3-0.6B~\cite{qwen3} as the LLM backbone and Utonia~\cite{zhang2026utonia} as the point encoder, for approximately 0.7B total parameters.
Training follows the two-stage recipe in \cref{sec:method:training}.
In Stage 1, we train only the projector for 3 epochs with a global batch size of 256 and a learning rate of $1\times10^{-3}$.
In Stage 2, we fine-tune the LLM and projector for 4 epochs while keeping the point encoder frozen, using a global batch size of 128, an LLM learning rate of $2\times10^{-5}$, a projector learning rate of $5\times10^{-4}$, and cosine scheduling with 5\% warmup.
We train models in bfloat16 on 8 NVIDIA H20 GPUs for about 7 hours.
At inference time, placing one object takes about one second on a single H20 GPU using roughly 3\,GB of memory, and the supplemental material breaks down the runtime of the full agent pipeline.

\subsection{Comparisons with State-of-the-Art Methods}
\label{sec:experiments:baselines}

\paragraph{Quantitative Comparisons}
\label{sec:experiments:baselines:quantitative} We compare external baselines using a single-object insertion protocol, where each method places one target object in a fixed scene.
ATISS \cite{atiss} and ReSpace \cite{bucher2025respace} serve as OBB-based references conditioned on object category.
Neither method accepts a specified placement instruction as input, so they are evaluated with only physical metrics.
For instruction conditioned baselines, we use a controlled LayoutVLM \cite{layoutvlm} adaptation: existing objects are fixed, only the target object is optimized, and the instruction is supplied as structured relations instead of being interpreted by a VLM, so that the baseline is scored on placement rather than on scene planning.
The supplemental material details this adaptation.
We also evaluate GPT-OSS-120B \cite{agarwal2025gpt} as a zero-shot large parameter model baseline.
For ATISS and ReSpace, which predict both pose and size, we scale the target point cloud by the predicted size before computing physical metrics.
\Cref{tab:sota-single-object} summarizes the quantitative results.

\input{sections/tabs/sota_single_object}

Our model achieves the strongest instruction-conditioned performance, with clear gains in both \textit{Within Valid Rate} and \textit{Instruction Valid Rate} over LayoutVLM and GPT-OSS-120B.
It also maintains competitive physical plausibility, with lower \textit{Floating Rate} and higher \textit{Support Ratio}.
ATISS obtains a lower \textit{Floating Rate} and a higher \textit{Support Ratio} because it tends to place objects on the floor instead of making fine-grained placements.
The supplemental material reports further comparisons under the same protocol, covering PlaceIt3D~\cite{placeit3d}, two substantially larger zero-shot LLMs, and an image-generation plus pose-estimation pipeline, none of which approaches our performance.

\input{sections/figs/ablation_cases}

\begin{table*}[!t]
  \begin{tabular}[t]{@{}p{0.48\linewidth}@{\hspace{0.04\linewidth}}p{0.48\linewidth}@{}}
    \input{sections/tabs/user_study}
    &
    \input{sections/tabs/ablation_pc}
  \end{tabular}
\end{table*}

\paragraph{Qualitative Comparisons}
\label{sec:experiments:qualitative} \Cref{fig:comparison-single} illustrates LayoutVLM, GPT-OSS-120B, and our method under the same single-object placement instructions.
LayoutVLM and GPT-OSS-120B respond to the instruction more directly, yet their predictions often miss fine-grained geometric constraints such as the target support surface, local free space, or containment region.
In contrast, our model uses point cloud to place the target object in geometrically valid regions.

\Cref{fig:qualitative-main} extends this qualitative comparison to whole-scene autoregressive inference.
Here the object set and the per-step instructions are taken from the held-out test scene, so the comparison isolates placement quality from planning, unlike the agent pipeline of \cref{sec:method:agent}, which derives both from a text prompt.
Across multi-step scene construction, our model produces more plausible and coherent layouts.
Additional inference results produced by our model on held-out test scenes are shown in \cref{fig:autoregressive-testset-gallery}.

\input{sections/figs/comparison_whole_scene.tex}
\input{sections/figs/testset_gallery.tex}
\input{sections/figs/agent_gallery.tex}

\paragraph{User Study}
\label{sec:experiments:user-study} \Cref{tab:user-study} presents the overall results of the user study.
We conduct a user study with 30 volunteers to evaluate single-object and whole-scene comparison results.
Each volunteer rates the displayed results on a five-point Likert scale, where 1 is the lowest quality and 5 is the highest quality.
For single-object placement results, the volunteers score physical plausibility and instruction following.
For whole-scene results, they score physical plausibility and visual aesthetics.
Since ATISS and ReSpace do not accept placement instructions, their single-object evaluations are omitted.

\subsection{Ablation Studies}
\label{sec:experiments:ablations}

We isolate the contribution of point cloud input, Stage 1 alignment pretraining, and high-quality training scenes through controlled ablations on the same test dataset.
The high-quality scenes include the Imaginarium and artist-crafted subsets.

\paragraph{Effect of point cloud input} The \textit{Instruction Valid Rate} remains comparable across variants, indicating point cloud input does not substantially improve generic instruction-conditioned validity.
Removing point cloud input weakens the model most directly on the \textit{Within Valid Rate}, where instruction following must be grounded in local containment geometry.
The same ablation also increases floating failures and reduces bottom-surface support, suggesting text-only context is insufficient for reliably grounding objects on available support surfaces.
\Cref{fig:case-study} corroborates these numbers: without point cloud input, the predicted objects frequently miss the referenced container, e.g., the fruit is placed outside the bookcase (column 1) even though the textual relation is parsed correctly.
Aggregate validity rates do not capture every fine-grained geometric failure. Measured directly, point cloud input reduces the target-center height error on \emph{within} samples by about 14\%, and a dedicated user study on this ablation prefers the point-cloud variant by a wide margin. Both analyses are reported in the supplemental material.

\paragraph{Effect of alignment pretraining} Removing Stage 1 alignment pretraining leads to a similar degradation, even though the model still receives point cloud input during placement fine-tuning.
The \textit{Within Valid Rate} decreases from 65.62\% to 62.69\%, showing the benefit of projector alignment.
\Cref{fig:case-study} exposes the same weakness qualitatively: without alignment pretraining, the fruit hovers in mid-air instead of resting on the shelf plank (column 1), and the blanket stays on the countertop and cuts into the basin instead of being placed inside the open compartment of the sink cabinet (column 4).
Together, these two ablations indicate explicit point cloud geometry and alignment pretraining are complementary.

\paragraph{Effect of high-quality scenes} Removing high-quality training scenes including the Imaginarium and artist-crafted scenes produces the largest drop in \textit{Within Valid Rate}.
This result suggests high-quality scenes primarily improve fine-grained geometric grounding for containment relations rather than generic instruction following validity.
The \textit{Floating Rate} and \textit{Support Ratio} also degrade, indicating these high-quality scenes provide useful supervision for stable object placement.
The qualitative cases in \cref{fig:case-study} mirror this trend: without high-quality scenes, the placed objects exhibit clear collisions or misalignment with the referenced anchors, e.g., the blanket cuts through the body of the sink cabinet and juts out beyond its side (column 4) and the briefcase overlaps the shelf frame (column 2) instead of resting inside it, whereas our full method produces collision-free placements that consistently satisfy the specified relations.
A stratified analysis in the supplementary material further shows this drop is concentrated on the high-quality test subset.

%% file: sections/figs/comparison_single_placement.tex
\begin{figure*}[!t]
  \centering
  \includegraphics[width=\textwidth]{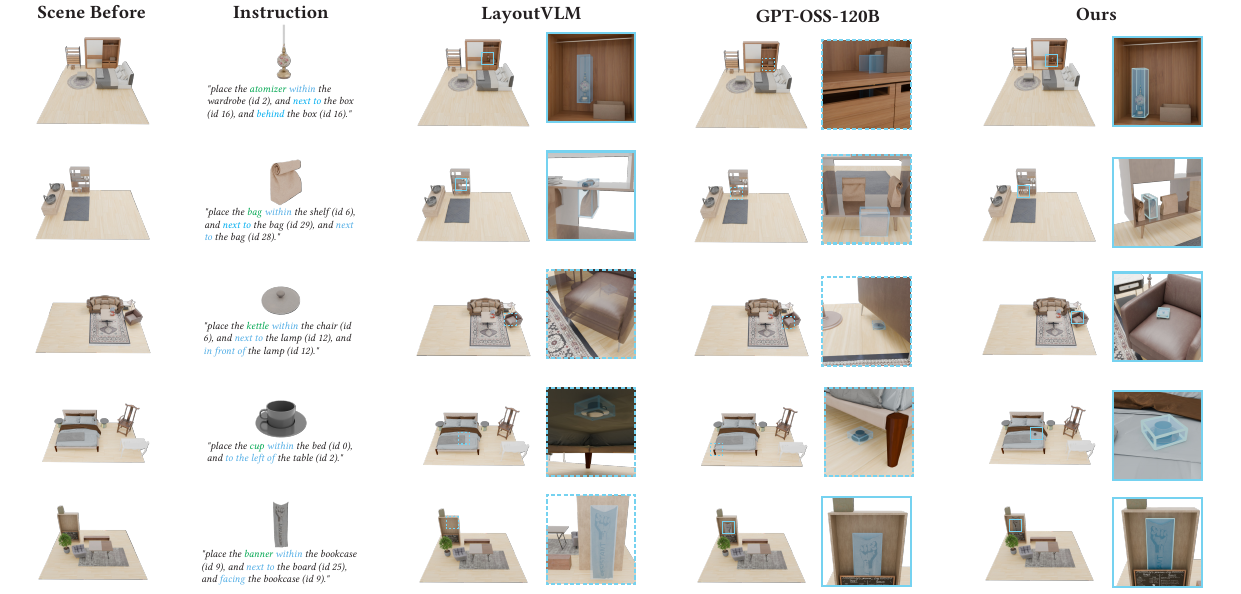}
  \caption{Qualitative comparisons of instruction-conditioned single-object placement results across instruction-aware baselines and our method.}
  \Description{Qualitative comparisons of instruction-conditioned single-object placement results across  baselines and our method.}
  \label{fig:comparison-single}
\end{figure*}

%% file: sections/tabs/sota_single_object.tex
\begin{table}[t]
  \centering
  \caption{Quantitative comparisons for single-object inference on the held-out test split. Cat., Instr., and PC denote category, instruction, and point cloud, respectively.
  Dashes indicate instruction-dependent metrics that are not applicable.}
  \label{tab:sota-single-object}
  \setlength{\tabcolsep}{1.5pt}
  \resizebox{\columnwidth}{!}{%
  \begin{tabular}{@{}llcccc@{}}
    \toprule
    Method & Input & Within. $\uparrow$ & Instruction. $\uparrow$ & Float. $\downarrow$ &
    Support. $\uparrow$ \\
    \midrule
    ATISS & OBB+Cat. & -- & -- & \textbf{0.93\%} & \textbf{99.20\%} \\
    ReSpace & OBB+Cat. & -- & -- & 5.79\% &
    94.20\% \\
    LayoutVLM & OBB+Instr. & 10.96\% & 37.71\% & 17.05\% & 82.91\% \\
    GPT-OSS-120B & OBB+Instr. & \underline{22.53\%} & \underline{42.29\%} & 7.64\% &
    93.54\% \\
    Ours (0.7B) & PC+Instr. & \textbf{65.62\%} & \textbf{78.59\%} & \underline{4.63\%} &
    \underline{96.03\%} \\
    \bottomrule
  \end{tabular}%
  }
\end{table}

%% file: sections/figs/ablation_cases.tex
\begin{figure*}[t]
  \centering
  \includegraphics[width=0.95\textwidth]{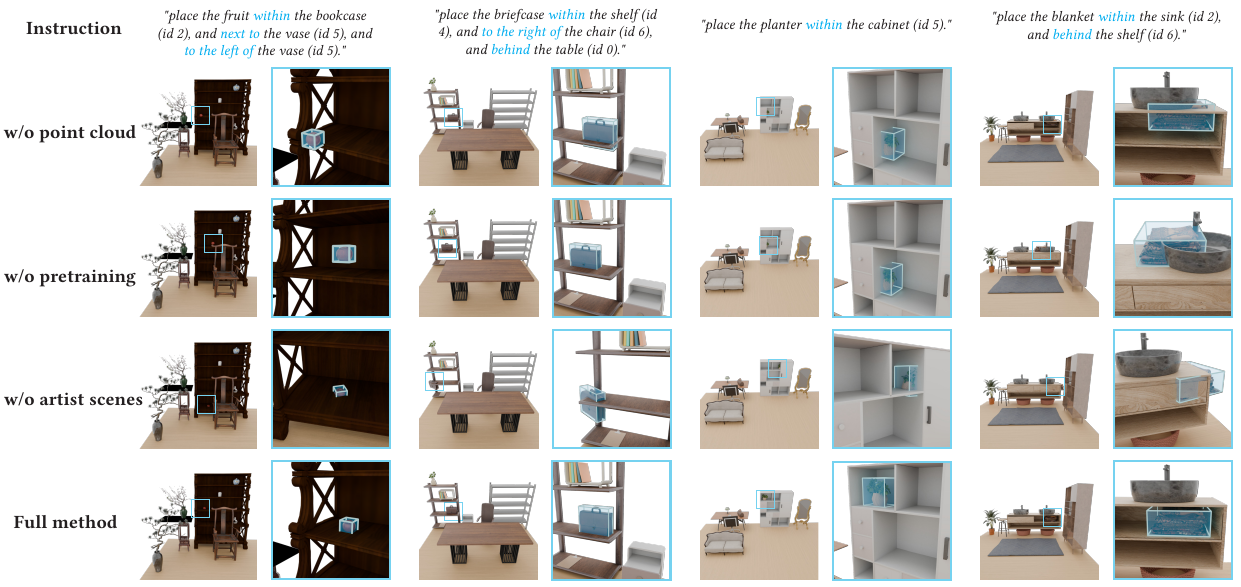}
  \caption{Qualitative ablation results on fine-grained placement. Each column shows one instruction, the predicted scene, and a zoomed crop around the target placement. Blue boxes mark inserted target objects.}
  \Description{Qualitative ablation results on fine-grained placement.}
  \label{fig:case-study}
\end{figure*}

%% file: sections/tabs/user_study.tex
\centering
\captionsetup{skip=0pt}
\caption{User-study scores from 30 participants. Dashes denote unavailable single-object scores for methods without instruction input.}
\label{tab:user-study}
\nointerlineskip
\vspace{4pt}
\footnotesize
\setlength{\tabcolsep}{2pt}
\resizebox{\linewidth}{!}{%
\begin{tabular}{@{}lcccc@{}}
  \toprule
  Method & \multicolumn{2}{c}{Single-object placement} &
  \multicolumn{2}{c}{Whole-scene result} \\
  \cmidrule(lr){2-3}\cmidrule(l){4-5}
  & Physical. $\uparrow$ & Instruction. $\uparrow$ &
  Physical. $\uparrow$ & Aesthetics. $\uparrow$ \\
  \midrule
  ATISS & -- & -- & 1.9 & 1.9 \\
  ReSpace & -- & -- & 2.7 & 2.9 \\
  LayoutVLM & 1.9 & 2.0 & \underline{3.0} & \underline{3.2} \\
  GPT-OSS-120B & \underline{2.7} & \underline{2.4} & 2.5 & 2.5 \\
  Ours (0.7B) & \textbf{4.5} & \textbf{4.5} & \textbf{3.9} & \textbf{4.1} \\
  \bottomrule
\end{tabular}%
}

%% file: sections/tabs/ablation_pc.tex
\centering
\captionsetup{skip=0pt}
\caption{Controlled ablations on the held-out test split. \emph{Within}/\emph{Instruction} report valid rates on containment/all samples. Arrows show preferred directions; bold/underlined values mark the best/second-best results.}
\label{tab:ablation-pc}
\nointerlineskip
\vspace{4pt}
\setlength{\tabcolsep}{2pt}
\begin{tabular}{@{}lcccc@{}}
  \toprule
  Method & Within. $\uparrow$ & Instruction. $\uparrow$ & Float. $\downarrow$ &
  Support. $\uparrow$ \\
  \midrule
  w/o point cloud & 62.40\% & \textbf{78.81\%} & 5.19\% & 95.35\% \\
  w/o pretraining & \underline{62.69\%} & 78.37\% & 5.20\% & 95.42\% \\
  w/o artist scenes & 60.32\% & 76.69\% & \underline{5.02\%} & \underline{95.88\%} \\
  Full method     & \textbf{65.62\%} & \underline{78.59\%} & \textbf{4.63\%} & \textbf{96.03\%} \\
  \bottomrule
\end{tabular}

%% file: sections/figs/comparison_whole_scene.tex
\begin{figure*}[!t]
  \centering
  \includegraphics[width=0.9\textwidth]{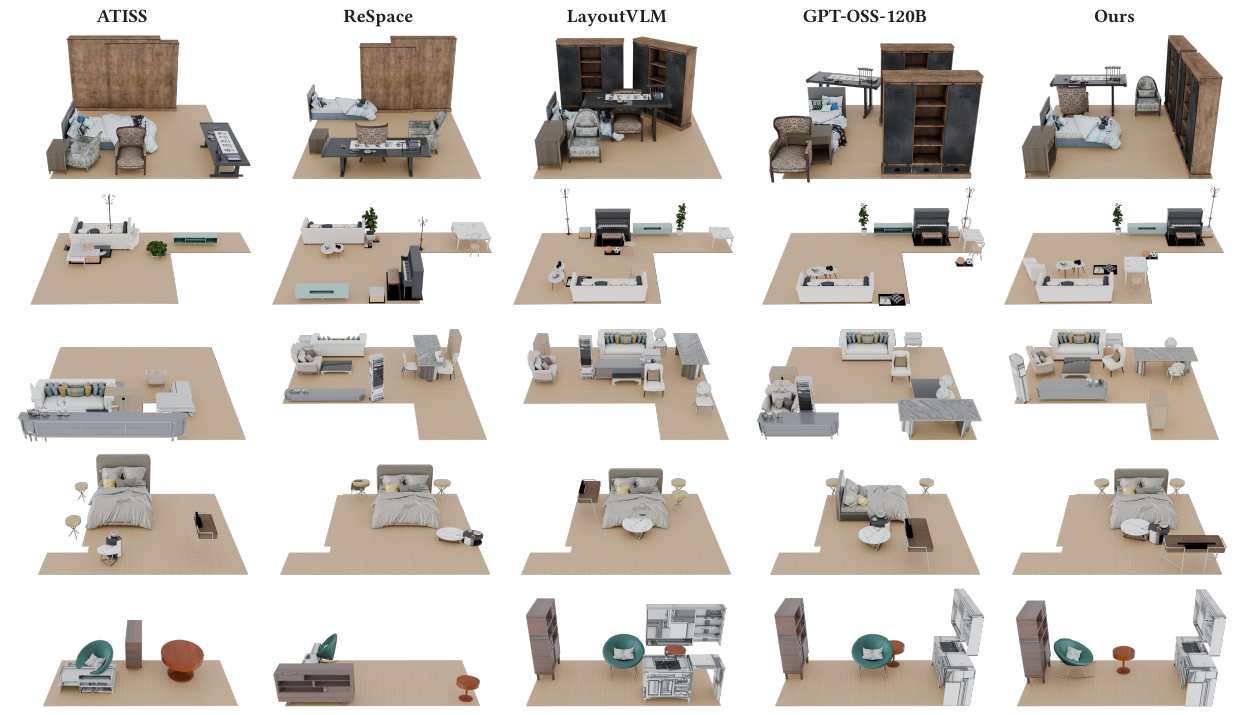}
  \caption{Qualitative comparison with baselines on whole-scene inference results. Each method receives the object set and the per-step placement instructions of a held-out test scene and places the objects one at a time, so the comparison isolates placement quality from planning.}
  \Description{Qualitative comparison with baselines on whole-scene inference results, where every method places the objects of a held-out test scene one at a time from the same per-step placement instructions.}
  \label{fig:qualitative-main}
\end{figure*}

%% file: sections/figs/testset_gallery.tex
\begin{figure*}[!p]
  \centering
  \includegraphics[width=0.88\textwidth]{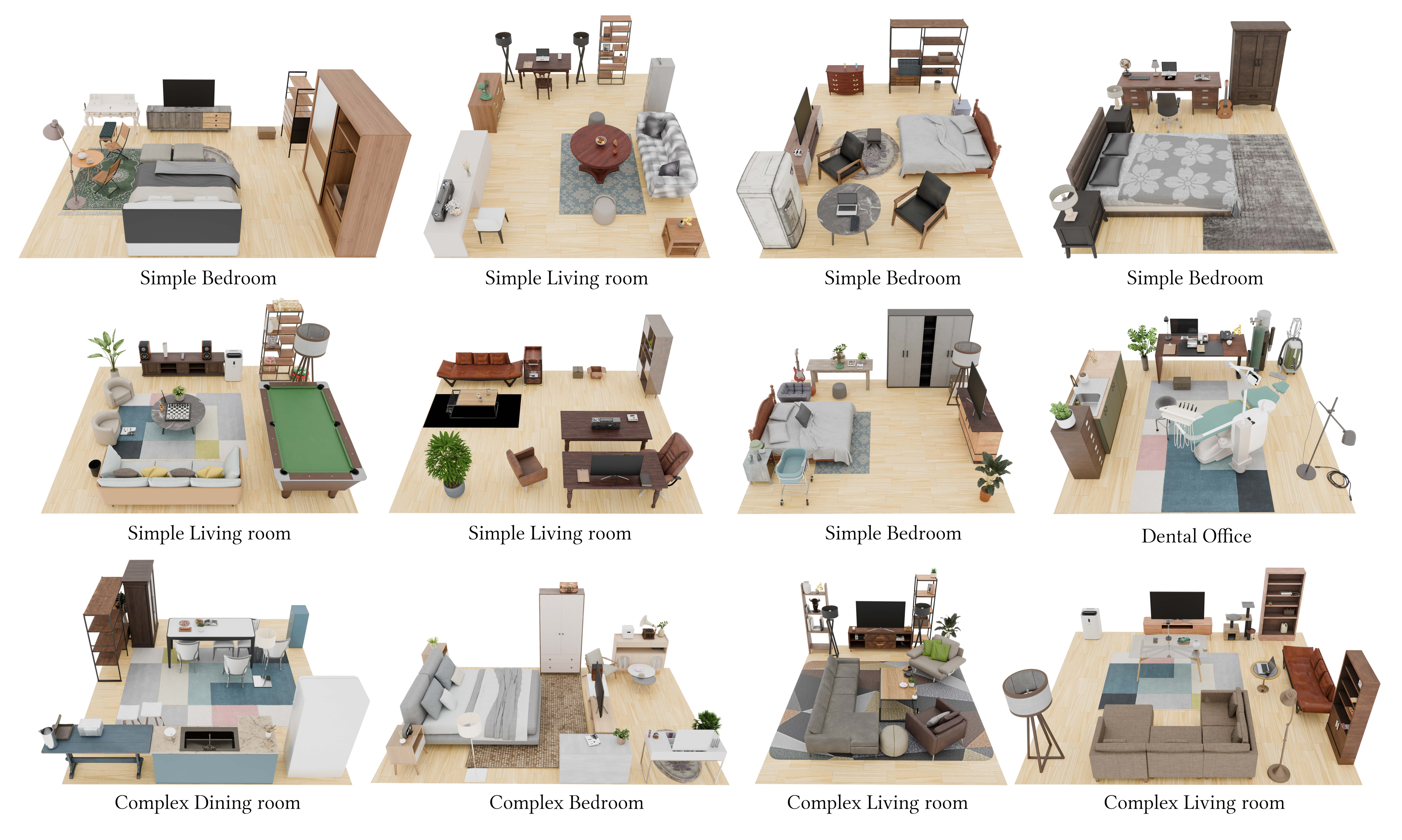}
  \caption{More autoregressive inference results of our model on held-out test scenes. Starting from scene-level object sets and placement instructions, our model sequentially predicts object poses and produces complete arrangements across different room types and layout complexity.}
  \Description{Gallery of twelve autoregressive inference results produced by our model on held-out test scenes, including bedrooms, living rooms, a dental office, and a dining room with varying layout complexity.}
  \label{fig:autoregressive-testset-gallery}
\end{figure*}

%% file: sections/figs/agent_gallery.tex
\begin{figure*}[!p]
  \centering
  \includegraphics[width=\textwidth]{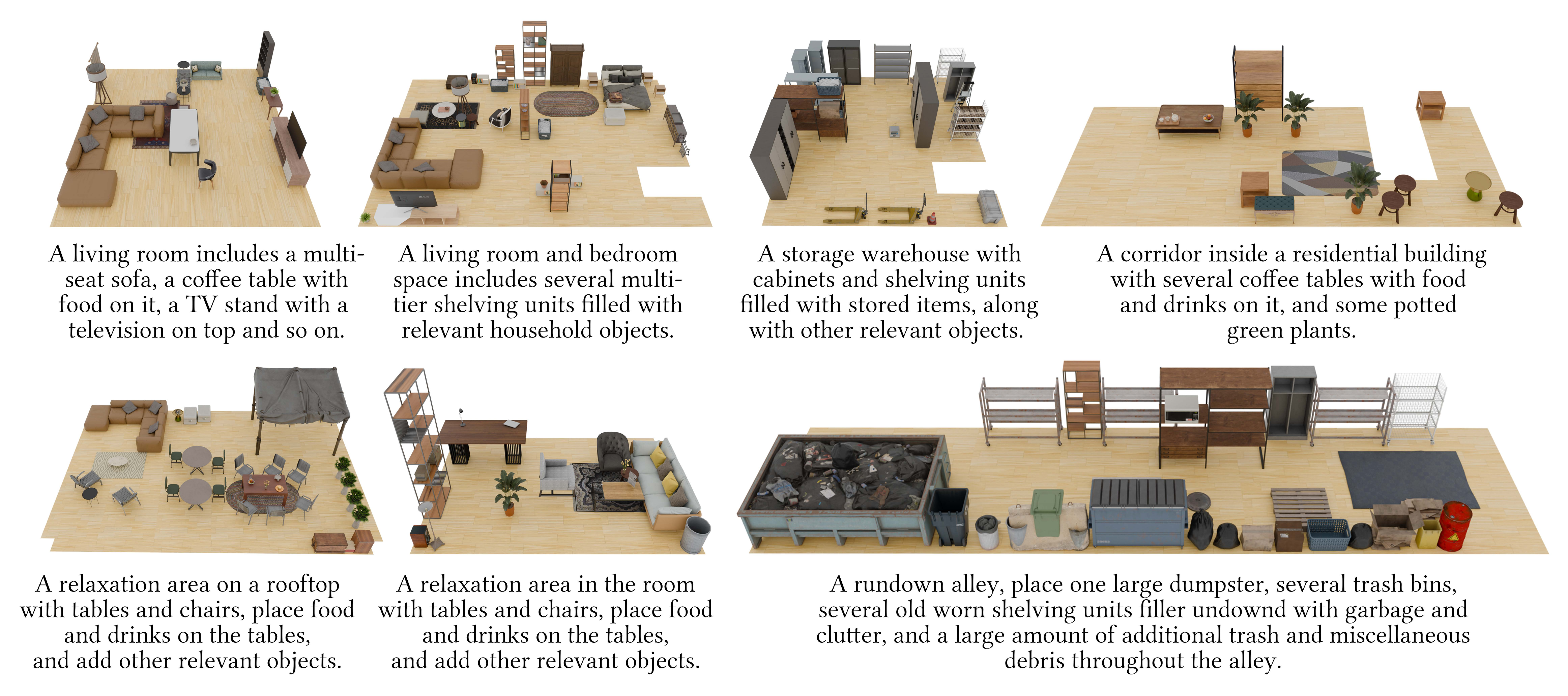}
  \caption{Gallery results from the agent pipeline for scene generation. Given text prompts and polygon floor boundary, the planner decomposes the request into object-level placement steps, then GALA predicts poses for the target objects to produce complete 3D scene arrangements.}
  \Description{Gallery of generated 3D scenes produced by the agent pipeline for scene generation from text prompts and polygon floor boundary.}
  \label{fig:agent-gallery}
\end{figure*}

%% file: sections/conclusions.tex
\section{Conclusions}
\label{sec:conclusion}

We propose GALA, a geometry-aware language model for controllable object arrangement.
By augmenting LLMs with point cloud perception and a two-stage training recipe, GALA can predict conditioned 6D poses.
Our approach captures fine-grained spatial constraints that are often missed by OBB-only methods.
We also integrate GALA seamlessly into an interactive agent framework for automated scene generation and editing.
Furthermore, we will release a dataset of high-fidelity scenes with verified, complex placements to facilitate future research.

\paragraph{Limitations} Despite its advantages, GALA currently supports only a predefined set of placement instructions and cannot yet handle open-ended placement requests. Its performance also degrades when generalizing to previously unseen container geometries. Moreover, because objects are placed autoregressively, errors in earlier placements may propagate to subsequent ones.

\paragraph{Future Work} Future research will extend this framework to handle noisy real-world scanned scenes, and support diverse interactive editing operations.
Another promising direction is explicit shape- or part-level supervision losses during two-stage training, which could strengthen geometric grounding.
Finally, because our model currently decodes pose tokens directly from context, future work could explore explicit chain-of-thought reasoning over geometry, in which the model first thinks about intermediate decisions, such as support-surface selection and the available space on that surface, before generating the pose.

%% file: sections/supplement_body.tex
\section{SpatialLM-PCG Dataset Expansion}
\label{app:spatiallm-pcg}

\input{sections/figs/spatiallm_pcg}
\input{sections/figs/dataset_gallery}

SpatialLM provides room-scale semantic layouts in the form of object categories, approximate 3D boxes, and coarse poses.
We use these layouts as structural scaffolds rather than final training scenes.
As shown in \cref{fig:spatiallm-pcg}, each proxy object is replaced by a plausible asset through category-constrained, size-and-shape-aware retrieval.
For a proxy object with category $c_i$ and box size
$\mathbf{s}_i=(w_i,d_i,h_i)$, we consider only assets with the same
semantic category. Each candidate asset $m$ has a canonical box size
$\mathbf{a}_m=(w_m,d_m,h_m)$ and is ranked by
\begin{equation}
  E(i,m)=
  \left\|\log \mathbf{a}_m-\log \mathbf{s}_i\right\|_1
  +\lambda
  \left\|\log \hat{\mathbf{a}}_m-\log \hat{\mathbf{s}}_i\right\|_1,
  \label{eq:pcg-asset-retrieval}
\end{equation}
where $\hat{\mathbf{s}}=\mathbf{s}/(wdh)^{1/3}$ is a scale-invariant shape descriptor, and $\lambda$ balances absolute size matching and proportion matching.
We select the asset minimizing \cref{eq:pcg-asset-retrieval}, then scale, orient, and place it to preserve the original layout.
The second term penalizes mismatched aspect ratios, reducing artifacts from extreme anisotropic scaling while keeping the spatial organization provided by SpatialLM.

After the major furniture layout is instantiated, we add small objects procedurally to enrich local interactions.
Candidate support and storage furniture, such as tables, cabinets, shelves, and TV stands, are assigned affordance regions derived from their geometry.
Small assets are sampled from category-compatible libraries and placed on, within, or near these regions with randomized offsets and orientations.
The generated objects remain anchored to plausible functional locations while increasing the frequency and diversity of support, containment, proximity, and directional relations in the placement instruction data.
The gallery in \cref{fig:dataset-gallery} illustrates the diversity of the curated scenes that result.

\section{Spatial Relation Definitions}
\label{app:relations}

We compute spatial relations from the posed oriented bounding boxes of the target object and each candidate reference object.
Each relation is represented as a discrete relation label together with the geometric quantities used to evaluate it.
These relation labels provide the intermediate supervision used to generate per-step natural-language instructions.
The full set of relation definitions is summarized in \cref{tab:relation-definitions}.

\paragraph{Notation}
Let $O_a$ denote the target object and $O_b$ a reference object. We use
$F_i$ for the XY footprint of object $O_i$, obtained by projecting its
oriented bounding box (OBB) vertices and taking the convex hull. Let
$d_{xy}(a,b)$ be the minimum distance between $F_a$ and $F_b$, and let
$z_i^{-}$ and $z_i^{+}$ denote the bottom and top of the posed 3D box.
The horizontal angle
$\theta_{ab}=\operatorname{atan2}(y_a-y_b,x_a-x_b)$ is measured in
degrees. The thresholds used in data generation are
$\tau_c=0.05\,\mathrm{m}$ for contact, $\tau_u=0.10\,\mathrm{m}$ for
under-object grounding, $\tau_n=0.50\,\mathrm{m}$ for next-to proximity,
$\tau_d=1.50\,\mathrm{m}$ for directional relations, $\eta_z=0.8$ for
horizontal height compatibility, $\eta_w=0.15$ for wall proximity,
$\eta_r=0.25$ for scene-region assignment, and $\tau_\theta=15^\circ$
for facing-angle alignment.

\input{sections/tabs/relation_definitions}

\section{Reference Object Selection}
\label{app:reference-selection}

Each autoregressive step conditions on the already placed objects.
To keep this context compact, we select a reference subset instead of passing the entire scene when the number of placed objects exceeds $K=8$.

\paragraph{Training-time selection}
During data generation, the target pose is known and can be used to identify geometrically relevant references.
We first compute all spatial relations between the target object and the existing objects.
Among these relations, we retain the most informative ones for the language instruction by sorting relation candidates by relation priority and then by footprint edge distance.
Support and containment relations (\texttt{on}, \texttt{under}, and \texttt{within}) have the highest priority, followed by next-to proximity and then directional and facing relations.
The objects explicitly mentioned by these selected relations are always retained.

In addition, if the target object has a parent in the dependency forest, we retain up to $S=6$ already placed siblings under the same parent, chosen by increasing footprint edge distance to the target.
These sibling references help represent local occupancy around shared supports or containers, such as multiple items on the same shelf or table.
The remaining slots are filled with the nearest existing objects by footprint edge distance.
The final selected set is sorted back into autoregressive order before being serialized.
If the mandatory references exceed the nominal cap, we keep all mandatory references rather than dropping supporting or sibling context.

\paragraph{Rollout-time selection}
At inference time, the target pose is unknown, so edge-distance ranking to the target is unavailable.
We therefore use an importance-based proxy.
Mandatory semantic dependencies specified by the planner are retained first; if a retained object is itself a sub-object, its parent is recursively retained as well.
The remaining slots are allocated between large anchors and recent context: 60\% of the available slots are assigned to objects with the largest footprint area, and the rest to the most recently introduced objects.
As in training, the selected references are returned in autoregressive order.
This policy preserves stable scene anchors while keeping the most recent local context available for the next placement.

\section{Design-Prior Object Ordering}
\label{app:object-ordering}

Our autoregressive training samples require a deterministic order for introducing objects into a scene.
Instead of relying on the arbitrary order stored in the source datasets, we use the design-prior heuristic of \cref{alg:design-prior-object-ordering}, which approximates a coarse-to-fine layout process.
It places large structural anchors first, then introduces smaller objects once the supporting or containing geometry they depend on is available.

We first infer a dependency forest from pairwise spatial relations.
If object $O_i$ is on, within, or under object $O_j$, then $O_j$ is treated as the parent of $O_i$, because $O_j$ provides the support, container, or overhead structure needed to localize $O_i$.
When multiple parent candidates exist, we prefer supporting relations over containment and under-object dependencies, and select the smallest parent footprint among candidates with the same priority.
This favors the most specific geometric reference, such as a shelf board over the entire cabinet.
Floor-covering objects are not used as supporting parents for \texttt{on} relations.

\begin{algorithm}[ht]
\SetAlgoLined
\KwIn{Object set $\mathcal{O}=\{O_i\}_{i=1}^{N}$}
\KwOut{Ordered object sequence $\pi$}
\BlankLine
\ForEach{$O_i \in \mathcal{O}$}{
  Compute footprint area $a_i$ and bottom height $b_i$\;
  Initialize $\mathrm{parent}(i) \leftarrow \varnothing$\;
}
\BlankLine
\ForEach{$O_i \in \mathcal{O}$}{
  $\mathcal{C}_i \leftarrow \varnothing$\tcp*{candidate parents}
  \ForEach{$O_j \in \mathcal{O}, j \neq i$}{
    Compute spatial relations between $O_i$ and $O_j$\;
    \If{$O_i$ is \texttt{on}, \texttt{within}, or \texttt{under} $O_j$}{
      Add $O_j$ to $\mathcal{C}_i$ with priority
      $\texttt{on}>\texttt{within}>\texttt{under}$\;
    }
  }
  \If{$\mathcal{C}_i \neq \varnothing$}{
    $\mathrm{parent}(i) \leftarrow$
    the candidate with highest priority and smallest footprint area\;
  }
}
\BlankLine
Break each cyclic dependency by making the lowest object in the cycle a root\;
$\mathcal{R} \leftarrow \{i \mid \mathrm{parent}(i)=\varnothing\}$\;
Sort $\mathcal{R}$ by increasing $(-a_i, b_i, i)$\;
$\pi \leftarrow \mathcal{R}$\;
$\mathcal{L} \leftarrow \mathcal{R}$\tcp*{current breadth-first layer}
\While{$\mathcal{L} \neq \varnothing$}{
  $\mathcal{L}' \leftarrow \varnothing$\;
  \ForEach{$p \in \mathcal{L}$ in order}{
    $\mathcal{K}_p \leftarrow \{i \mid \mathrm{parent}(i)=p\}$\;
    Sort $\mathcal{K}_p$ by increasing $(-a_i, b_i, i)$\;
    Append $\mathcal{K}_p$ to $\mathcal{L}'$\;
  }
  Append $\mathcal{L}'$ to $\pi$\;
  $\mathcal{L} \leftarrow \mathcal{L}'$\;
}
\Return{$\pi$}\;
\caption{Design-prior object ordering.}
\label{alg:design-prior-object-ordering}
\end{algorithm}

The resulting sequence is not meant to model a single universal human ordering.
Instead, it encodes the practical dependency: fine-grained objects are easier to place after their geometric anchors are present.
In the generated training data, object identifiers are reassigned according to this sequence, so each autoregressive step conditions only on objects already introduced by the design-prior order.

\section{Evaluation Metric Definitions}
\label{app:metric-definitions}

This section gives the exact definitions of the metrics reported in the main paper and in this supplement.
Each evaluation sample contains an existing scene, a target object point cloud, a predicted 6D pose, and a set of expected instruction relations.
All methods, including the baselines, are scored by the same evaluator.

\paragraph{Collision}
Collisions are detected directly on raw point clouds, without convex-hull or mesh conversion, so that hollow containers such as cabinets and shelves are treated as the open geometry they are.
We voxelize the posed target point cloud and every existing-object point cloud at a resolution of $0.02\,\mathrm{m}$, and denote by $V_{\mathrm{tar}}$ the set of occupied target voxels and by $V_{\mathrm{ovl}}$ the subset of them also occupied by some existing object.
A placement is marked as colliding when
\begin{equation}
  |V_{\mathrm{ovl}}| \,/\, |V_{\mathrm{tar}}| > 0.20 .
\end{equation}
Thresholding a ratio rather than an absolute overlap keeps the criterion scale-invariant across object sizes and tolerates the thin-shell interpenetration that point-cloud discretization inevitably produces, while still rejecting placements that pass through existing geometry.

\paragraph{Floating Rate and Support Ratio}
To measure physical support, we first transform the target object point cloud by the predicted pose and take the lowest 10\% of its points by $z$ value as the bottom surface $B$.
A 2D support height map is built from the floor point cloud and from existing-object points no higher than the target bottom plus a small gap tolerance.
For each bottom point $p\in B$, we search a $5\times5$ voxel neighborhood in this map and record the nearest supporting height below the point.
The point is considered supported if its vertical gap to this support surface is at most $\tau_{\mathrm{gap}}=0.05\,\mathrm{m}$.
The \textit{Support Ratio} is
\begin{equation}
  r_{\mathrm{sup}}
  =
  \frac{1}{|B|}
  \sum_{p\in B}
  \mathbb{I}
  \left[
    \operatorname{gap}(p)\leq \tau_{\mathrm{gap}}
  \right].
\end{equation}
If the target object height is available and the predicted center height is within $0.05\,\mathrm{m}$ of the floor-contact height from both sides, the sample is treated as floor-supported directly.
For the full point-cloud check, we also flag a placement as sinking when the median bottom-surface gap is below $-\tau_{\mathrm{gap}}$, indicating the object has sunk into the supporting surface.
A placement is counted as floating when $r_{\mathrm{sup}}<0.30$ or when this sinking condition holds.
We report \textit{Floating Rate} as the fraction of floating placements and \textit{Support Ratio} as the mean of $r_{\mathrm{sup}}$, where samples resolved by the floor-contact shortcut above contribute $r_{\mathrm{sup}}=1$.

\paragraph{Instruction Valid Rate and Within Valid Rate}
For each sample $i$, let $P_i$ indicate whether the model output can be parsed as a valid 6D pose, $S_i$ indicate whether all expected instruction relations are satisfied, $C_i$ indicate collision, and $F_i$ indicate floating.
Let $I_i$ be the \textit{Instruction Valid} indicator:
\begin{equation}
  I_i
  =
  P_i \wedge S_i \wedge \neg C_i \wedge \neg F_i .
\end{equation}
The \textit{Instruction Valid Rate} is the average of this indicator over all evaluated samples.
The \textit{Within Valid Rate} uses the same indicator on the subset $\mathcal{W}$ of samples whose expected relations contain at least one \texttt{within} relation:
\begin{equation}
  \operatorname{WithinValid}
  =
  \frac{1}{|\mathcal{W}|}
  \sum_{i\in\mathcal{W}}
  I_i .
\end{equation}
This subset is particularly sensitive to point-level geometry because containment and insertion often depend on openings, shelves, and support surfaces missing from coarse bounding boxes.

\section{Checkpoint Selection}
\label{app:checkpoint-selection}

\input{sections/figs/validation_loss}

For the internal comparison in the main paper, we selected checkpoints by validation loss before inspecting test-set metrics.
As shown in \cref{fig:validation-loss}, both the text-only and point-cloud variants reached their lowest validation loss at step 6240 (epoch 2.67), with validation losses of 0.341 and 0.332, respectively.
We therefore report test metrics using this validation-selected checkpoint rather than the final checkpoint after four epochs.

\section{LayoutVLM Adaptation}
\label{app:layoutvlm-adaptation}

LayoutVLM synthesizes a scene by having a VLM propose differentiable spatial constraints and a gradient-based solver optimize the asset poses jointly.
Since our protocol fixes the scene and inserts a single object, we supply the instruction directly as the structured relations of \cref{tab:relation-definitions}, which grants the baseline oracle instruction parsing, and let its solver optimize only the target pose while all existing objects stay frozen.
Like the original method, the adaptation reasons over oriented bounding boxes and derives the placement height from the reference box, using its top face for \texttt{on} and its center height for \texttt{within}.
For the whole-scene comparison in the main paper, it runs autoregressively under the same protocol as our model, inserting objects one at a time in the order and with the per-step instructions of the held-out test scene.

\section{Additional Baseline Comparisons}
\label{app:additional-baselines}

\Cref{tab:additional-baselines} evaluates additional methods using the same single-object insertion protocol and geometric metrics as in the main paper.
All existing scene objects and the target geometry remain fixed, so each method is evaluated only on the predicted placement.

\begin{table}[H]
  \centering
  \caption{Additional single-object placement baselines on the full held-out test split.}
  \label{tab:additional-baselines}
  \setlength{\tabcolsep}{2pt}
  \resizebox{\columnwidth}{!}{%
  \begin{tabular}{@{}lcccc@{}}
    \toprule
    Method & Within. $\uparrow$ & Instruction. $\uparrow$ & Floating. $\downarrow$ & Support. $\uparrow$ \\
    \midrule
    PlaceIt3D & 0.32\% & 1.39\% & 67.53\% & 35.96\% \\
    DeepSeek-V4-Flash & 11.53\% & 30.02\% & 17.89\% & 84.75\% \\
    Qwen3.5-397B-A17B & 17.35\% & 37.28\% & 12.87\% & 89.11\% \\
    Ours (0.7B) & \textbf{65.62\%} & \textbf{78.59\%} & \textbf{4.63\%} & \textbf{96.03\%} \\
    \bottomrule
  \end{tabular}%
  }
\end{table}

\paragraph{PlaceIt3D adaptation}
PlaceIt3D predicts a spatial placement mask rather than a numeric 6D pose.
We adapt its selected placement to the target-pose representation required by our evaluator while retaining the original target geometry and fixed scene context, and then apply exactly the same relation and physical-validity checks used for all methods.

\paragraph{Stronger zero-shot LLMs}
DeepSeek-V4-Flash and Qwen3.5-397B-A17B receive the same serialized scene metadata and placement instruction used by our zero-shot LLM baseline, without point-cloud tokens or task-specific fine-tuning.
Although both models are substantially larger than GALA, their results show that model scale alone does not provide reliable metric 6D placement.

\paragraph{Image-generation and pose-estimation pipeline}
We additionally evaluate Nano-Banana + SAM + FoundationPose on 100 randomly sampled test cases.
For each case, we render four input views, generate an image containing the target object, segment it, recover its pose, and conservatively retain the best recovered pose across the four views under our evaluator.
\Cref{tab:image-pipeline-baseline} therefore reports this baseline under a favorable multi-view best-pose protocol and compares it with GALA on exactly the same 100 cases.

\begin{table}[H]
  \centering
  \caption{Image-pipeline comparison on the same 100-case subset using four views per case and the best recovered pose. The first row abbreviates the Nano-Banana + SAM + FoundationPose pipeline.}
  \label{tab:image-pipeline-baseline}
  \setlength{\tabcolsep}{2pt}
  \resizebox{\columnwidth}{!}{%
  \begin{tabular}{@{}lcccc@{}}
    \toprule
    Method & Within. $\uparrow$ & Instruction. $\uparrow$ & Floating. $\downarrow$ & Support. $\uparrow$ \\
    \midrule
    Nano-Banana + FoundationPose & 12.82\% & 15.00\% & 69.00\% & 31.21\% \\
    Ours & \textbf{66.67\%} & \textbf{79.00\%} & \textbf{6.00\%} & \textbf{92.02\%} \\
    \bottomrule
  \end{tabular}%
  }
\end{table}

This image-based pipeline is view-dependent, can alter the existing scene during image generation, and accumulates error across synthesis, segmentation, and pose recovery.
It also takes approximately one minute per object, compared with approximately one second for direct GALA placement.

\section{Stratified Ablation on High-quality Scenes}
\label{app:artist-crafted-ablation}

\input{sections/tabs/artist_quality_ablation}

\Cref{tab:artist-crafted-subset-ablation} shows that removing high-quality training scenes primarily affects the high-quality test subset, which contains Imaginarium and artist-crafted scenes.
The \textit{Within Valid Rate} drops from 52.82\% to 39.47\% on this subset, compared with a smaller drop from 68.65\% to 65.44\% on the 3D-FRONT and SpatialLM-PCG subset.
\textit{Instruction Valid Rate} shows the same concentration, and the \textit{Floating Rate} increase is confined to the high-quality subset, suggesting high-quality scenes provide useful supervision for designer-curated layouts.

\section{Fine-Grained Benefits of Point Clouds}
\label{app:point-cloud-benefits}

\paragraph{Within-sample height localization}
Containment often requires selecting a particular shelf or compartment height that cannot be recovered from the container's overall bounding box.
On all 3,165 \texttt{within} samples, we therefore measure the target-center height error
\begin{equation}
  \Delta z_i = \left|z_i^{\mathrm{pred}} - z_i^{\mathrm{gt}}\right|.
  \label{eq:within-height-error}
\end{equation}
Averaged over these samples, \cref{eq:within-height-error} is $0.0747\,\mathrm{m}$ with point clouds and $0.0868\,\mathrm{m}$ without point clouds, a reduction of approximately 14\%.
A paired Wilcoxon test gives $p=6.7\times10^{-20}$, showing that this difference is statistically significant.

\begin{table}[H]
  \centering
  \caption{Point-cloud analysis on \texttt{within} samples and in a separate ablation user study.}
  \label{tab:point-cloud-additional}
  \setlength{\tabcolsep}{5pt}
  \begin{tabular}{@{}lcc@{}}
    \toprule
    Metric & w/ point cloud & w/o point cloud \\
    \midrule
    Height error (m) $\downarrow$ & \textbf{0.0747} & 0.0868 \\
    User-study score (1--5) $\uparrow$ & \textbf{4.14} & 1.91 \\
    \bottomrule
  \end{tabular}
\end{table}

\paragraph{Text-only ablation user study}
In a study separate from the main-paper comparison, 24 participants rated 23 single-object placement results produced by the point-cloud and text-only variants on a five-point scale.
As shown in \cref{tab:point-cloud-additional}, the point-cloud model receives a mean score of 4.14, compared with 1.91 for the text-only model.
This larger perceptual gap complements the aggregate validity metrics, which do not capture every fine-grained geometric failure.

\paragraph{L-shaped sink counter case}
The lower-left result of Fig.~9 in the main paper, labeled \emph{Complex Dining room}, contains an L-shaped sink counter whose top combines a recessed basin with an unoccupied corner.
The predicted placement respects the usable local surface while avoiding both regions, illustrating a shape-dependent decision that an OBB alone cannot express.

\section{Generalization Analysis on Unseen Assets}
\label{app:unseen-assets}

The scene-level split already holds out complete room instances.
To test asset-level generalization more directly, we isolate 128 \texttt{within} samples whose reference containers comprise eight asset instances that never appear in training, including bookcases, shelves, islands, chairs, TV stands, and vanities.
\Cref{tab:unseen-assets} compares these samples with a size-matched set using seen container instances.

\begin{table}[H]
  \centering
  \caption{Asset-level generalization on two size-matched sets of 128 \texttt{within}-relation samples.}
  \label{tab:unseen-assets}
  \setlength{\tabcolsep}{3pt}
  \resizebox{\columnwidth}{!}{%
  \begin{tabular}{@{}lccc@{}}
    \toprule
    Group & Within. $\uparrow$ & Floating. $\downarrow$ & Support. $\uparrow$ \\
    \midrule
    Seen container, w/ point cloud & 61.72\% & 16.41\% & 84.21\% \\
    Unseen container, w/ point cloud & 59.38\% & 31.25\% & 68.34\% \\
    \bottomrule
  \end{tabular}%
  }
\end{table}

Containment accuracy degrades on unfamiliar containers, but only mildly: the \textit{Within Valid Rate} falls by 2.34 percentage points relative to the seen comparison.
Physical plausibility degrades considerably more, with the \textit{Floating Rate} roughly doubling and the \textit{Support Ratio} dropping by about 16 points, so these results show partial rather than equal robustness on unseen container instances.
Both groups consist exclusively of \texttt{within} samples, whose targets must reach an interior support surface rather than the floor, so the rates in this table are structurally higher than aggregate rates over all relation types and are meaningful only relative to each other.
This analysis is limited to eight assets and 128 samples from a concentrated set of furniture types, and it evaluates generalization to unseen instances within known categories rather than to unseen categories.

\section{Runtime Analysis}
\label{app:runtime}

\begin{table}[H]
  \centering
  \caption{Runtime breakdown of the end-to-end agent pipeline. API timings depend on the remote service, whereas GALA placement is measured locally.}
  \label{tab:runtime}
  \setlength{\tabcolsep}{2pt}
  \resizebox{\columnwidth}{!}{%
  \begin{tabular}{@{}lllll@{}}
    \toprule
    Component & Frequency & Runtime & Execution & Local memory/hardware \\
    \midrule
    Scene planner & Once per scene & $\sim$100 s & API & -- \\
    Instruction generator & Once per object & $\sim$70 s & API & -- \\
    GALA placement & Once per object & $\sim$1 s & Local inference & $\sim$3 GB, 1 NVIDIA H20 \\
    \bottomrule
  \end{tabular}%
  }
\end{table}

As \cref{tab:runtime} shows, the two upstream LLM calls dominate end-to-end latency.
GALA itself is lightweight at approximately one second per object on an H20 GPU, so faster planning and instruction-generation endpoints or local deployment can reduce total runtime without changing the placement model.

\clearpage
\onecolumn

\section{Prompt Templates}
\label{app:prompts}

The agent pipeline uses two system prompts, reproduced below: \cref{lst:scene-planning-prompt} drives the scene planner, and \cref{lst:instruction-generation-prompt} drives the per-object instruction generator.

\begin{lstlisting}[style=promptbox,caption={Rendered system prompt for scene planning.},label={lst:scene-planning-prompt}]
You are a professional interior scene planner. Your task is to analyze user requirements and create a detailed scene plan.

## Your Task
-  Understand user's scene requirements
- Expand and refine the scene description with professional knowledge
- Generate a complete object list with appropriate sizes and placement order
- Consider spatial relationships and functional requirements

## Guidelines
- Objects should be ordered by priority (larger/anchor/bottom objects first)
- Estimated sizes should be in meters with two decimal places (e.g., [1.20, 0.60, 0.75]) and realistic for the object type. For beds, the estimated height should include the headboard height, not just the mattress height. A typical bed with headboard is 1.0-1.4 meters tall.
- Include necessary furniture based on room type (e.g., bed for bedroom)
- Consider traffic flow and functionality
- The 'description' field should be specific enough for asset retrieval (include material, color, style)
- Only include floor-standing furniture and objects that can be placed on/within/under other objects.
- Do NOT include: ceiling-mounted or suspended items of any kind, especially anything hanging from the ceiling such as chandeliers, pendant lights, ceiling lamps, ceiling fans with lights, or any other overhead lighting fixtures.
- Do NOT include: wall-mounted items (mirrors, shelves, paintings) or any lighting fixtures. If lighting is needed, rely on floor lamps or table lamps only.
- Use "depends_on" to specify which earlier objects this object's placement depends on (by their IDs). This includes: objects it will be placed ON, WITHIN, or UNDER; objects it must be placed NEXT TO or FACING. Standalone objects with no spatial dependency should omit this field or use an empty list.

## Output Format
You must output a valid JSON object with the following structure:
{
    "thinking": "Your step-by-step reasoning about the scene design...",
    "scene_type": "bedroom/livingroom/diningroom/office/...",
    "scene_description": "A detailed description of the target scene",
    "objects": [
        {
            "id": 0,
            "category": "bed/sofa/table/chair/...",
            "description": "detailed description for asset retrieval",
            "estimated_size": [width, depth, height]
        },
        {
            "id": 1,
            "category": "nightstand",
            "description": "...",
            "estimated_size": [0.50, 0.40, 0.55],
            "depends_on": [0]
        },
        ...
    ]
}
\end{lstlisting}

\begin{lstlisting}[style=promptbox,caption={Compact rendering of the system prompt for instruction generation.},label={lst:instruction-generation-prompt}]
You are a professional scene layout assistant. Your task is to generate placement instructions for objects in a scene.

## Coordinate and Relation Vocabulary
- Use a right-handed coordinate system: +X = right/east, +Y = front/north, +Z = up.
- Absolute relations:
  * Wall: "against east/west/north/south wall"
  * Corner: "at northeast/northwest/southeast/southwest corner"
  * Region: "in front/back/left/right/center region"
- Relative relations:
  * Directional: "left", "right", "in_front", "behind" (edge distance < 1.5m)
  * Proximity/orientation: "next_to" (edge distance < 0.5m), "face_to"
  * Support/containment: "on", "under", "within"

## CRITICAL: Space Awareness Before Using Direction Relations
Before using a direction relation, compute the object's bounding box and check that the target direction is not blocked by a wall or boundary. If an object is already against a wall on one side, do NOT place another object further in that direction.

## Guidelines
- First analyze the scene context and existing objects to understand the scene layout
- Calculate available space in each direction before choosing a relation
- Use existing object IDs and short descriptions as references
- Combine absolute and relative relations when appropriate
- Be specific but concise
- Consider the object's function and typical placement patterns

## Main Objects vs Sub-Objects
- Main objects are standalone room-level objects (e.g., bed, sofa, desk, wardrobe). Set "parent_id": null and include appropriate absolute relations.
- Sub-objects are placed ON/WITHIN/UNDER a parent object (e.g., pillow on a bed, book on a shelf, chair under a desk). Set "parent_id" to the parent ID and omit absolute relations.

## Relation Rules
- Include at most 3 relations total, prioritized as: on/within/under > next_to > directional.
- Use at most one wall relation; use a corner relation instead of combining two wall relations.
- For any sub-object resting on or contacting a bed or sofa, use "within" rather than "on".
- If a parent already has sub-objects, add a directional relation to an existing sibling to avoid collision.

## Output Format
{
    "thinking": "Your step-by-step reasoning about where to place the object...",
    "parent_id": null,
    "expected_relations": {
        "relative_relations": [
            {"type": "relation_type", "target_id": reference_object_id}
        ],
        "absolute_relations": ["relation_type"]
    }
}

## Representative Examples
- Main object: next_to + directional relation to an existing anchor, with an optional wall relation.
- Sub-object: support/containment relation to its parent only.
- Crowded parent: support/containment relation to the parent plus a directional relation to an existing sibling.
\end{lstlisting}

%% file: sections/figs/spatiallm_pcg.tex
\begin{figure*}[t]
  \centering
  \includegraphics[width=\textwidth]{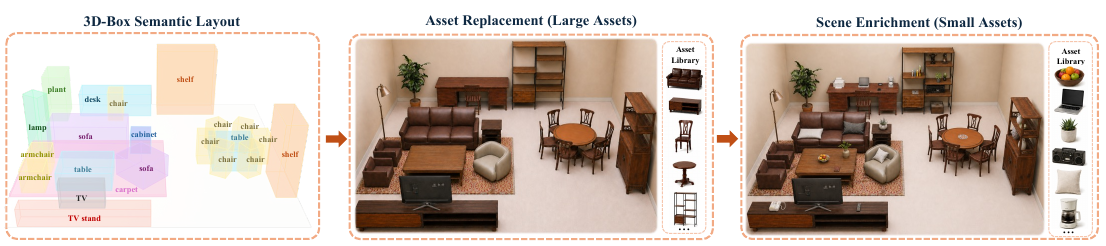}
  \caption{Procedural expansion of SpatialLM layouts. Starting from a 3D-box semantic layout, we replace large proxy objects with compatible assets from a curated library and then enrich the scene with small objects placed according to the geometry and affordances of the instantiated furniture.}
  \Description{Pipeline for SpatialLM-PCG construction, showing an input
  3D bounding-box semantic layout, replacement of large furniture assets,
  and enrichment with small objects on or near furniture.}
  \label{fig:spatiallm-pcg}
\end{figure*}

%% file: sections/figs/dataset_gallery.tex
\begin{figure*}[!t]
  \centering
  \includegraphics[width=\textwidth]{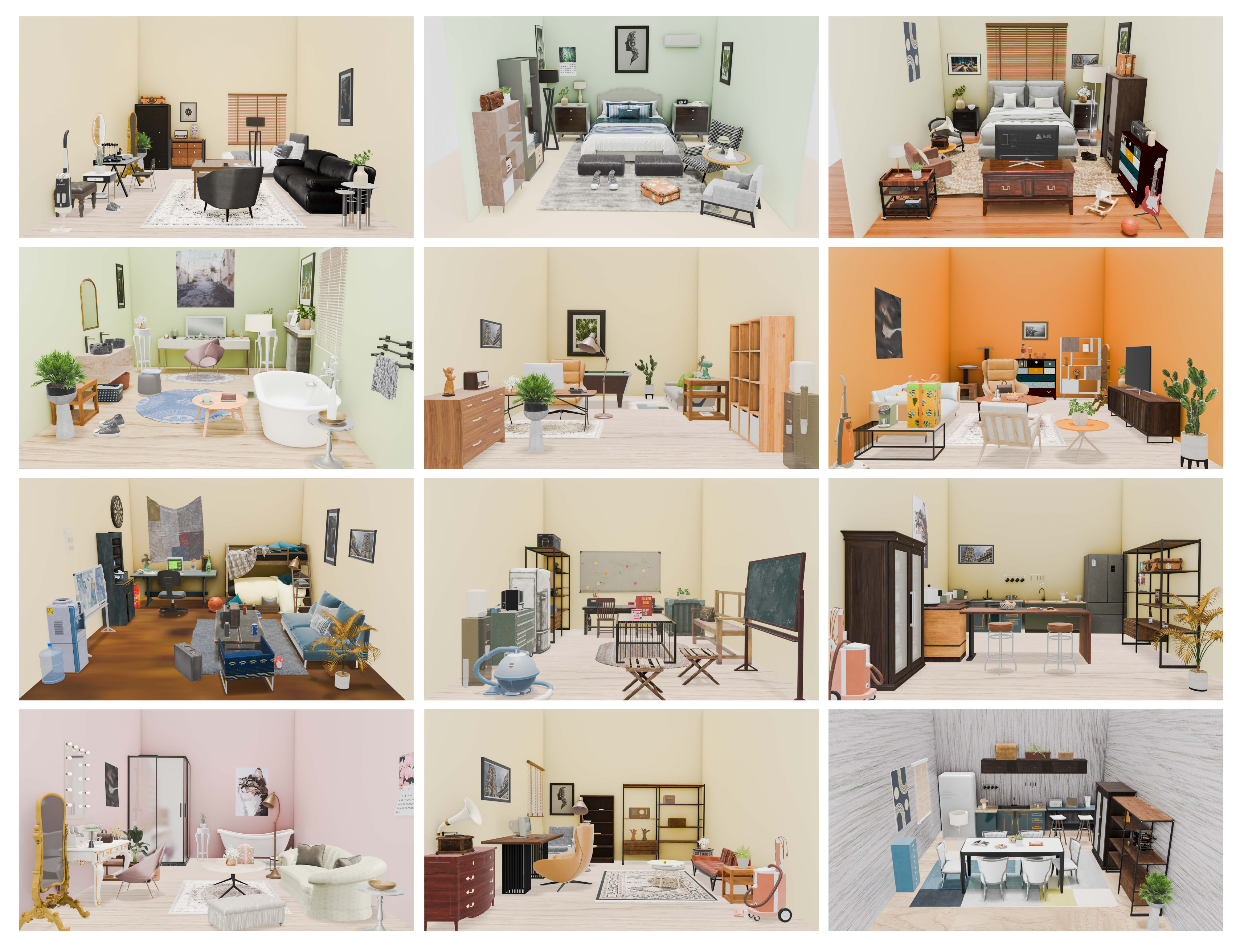}
  \caption{Overview of the hand-crafted scene dataset used for geometry-aware object arrangement. The gallery illustrates the diversity of room types, object categories, and fine-grained support and containment configurations.}
  \Description{Dataset gallery showing diverse indoor scenes with varied room types, furniture categories, and fine-grained object arrangements.}
  \label{fig:dataset-gallery}
\end{figure*}

%% file: sections/tabs/relation_definitions.tex
\begin{table*}[t]
  \centering
  \small
  \caption{Definitions of spatial relations used for instruction
  generation.}
  \label{tab:relation-definitions}
  \begin{tabular}{@{}p{0.16\textwidth}p{0.18\textwidth}p{0.48\textwidth}p{0.12\textwidth}@{}}
    \toprule
    Relation type & Relation & Geometric condition & Thresholds \\
    \midrule
    Scene boundary &
    \begin{tabular}[t]{@{}l@{}}
      \texttt{against east wall}\\
      \texttt{against west wall}\\
      \texttt{against north wall}\\
      \texttt{against south wall}
    \end{tabular} &
    The object is floor-supported ($z_a^{-}\leq\tau_u$). The distance
    from $F_a$ to the corresponding boundary edge is below
    $\eta_w\,\mathrm{diag}_{xy}(O_a)$. Directional wall names are assigned
    from the edge outward normal. &
    $\tau_u,\eta_w$ \\
    Scene boundary &
    \begin{tabular}[t]{@{}l@{}}
      \texttt{at northeast corner}\\
      \texttt{at northwest corner}\\
      \texttt{at southeast corner}\\
      \texttt{at southwest corner}
    \end{tabular} &
    The object is floor-supported and its footprint is within the wall
    threshold of two adjacent convex boundary edges. The corner name is
    determined from the two outward normals, with a centroid-based quadrant
    fallback for irregular boundaries. &
    $\tau_u,\eta_w$ \\
    Scene region &
    \begin{tabular}[t]{@{}l@{}}
      \texttt{in front region}\\
      \texttt{in back region}\\
      \texttt{in left region}\\
      \texttt{in right region}\\
      \texttt{in center region}
    \end{tabular} &
    Used when no wall or corner relation is emitted. The object center is
    compared with the scene AABB center: front/back use the sign of the
    Y offset and left/right use the sign of the X offset. A directional
    region requires the corresponding offset magnitude to exceed
    $\eta_r$ times the scene depth or width; otherwise the object is in the
    center region. &
    $\eta_r$ \\
    Horizontal proximity &
    \texttt{next to} &
    The footprint distance satisfies $d_{xy}(a,b)<\tau_n$. Horizontal
    relations are suppressed when an \texttt{on}, \texttt{under}, or
    \texttt{within} relation already holds, and require compatible bottom
    heights:
    $|z_a^{-}-z_b^{-}|\leq \eta_z \max(h_a,h_b)$, where
    $h_i=z_i^{+}-z_i^{-}$. &
    $\tau_n,\eta_z$ \\
    Horizontal direction &
    \begin{tabular}[t]{@{}l@{}}
      \texttt{left}\\
      \texttt{right}\\
      \texttt{in front}\\
      \texttt{behind}
    \end{tabular} &
    The footprint distance satisfies $d_{xy}(a,b)<\tau_d$ and the same
    height and vertical-relation exclusions as above hold. The relation is
    selected by $\theta_{ab}$:
    left for $[135^\circ,180^\circ]\cup(-180^\circ,-135^\circ]$,
    right for $[-45^\circ,45^\circ)$, in front for
    $[45^\circ,135^\circ)$, and behind for $[-135^\circ,-45^\circ)$. &
    $\tau_d,\eta_z$ \\
    Support &
    \texttt{on} &
    The target center projected to XY lies inside $F_b$, and the target
    bottom is close to the reference top:
    $|z_a^{-}-z_b^{+}|<\tau_c$. &
    $\tau_c$ \\
    Containment &
    \texttt{within} &
    All eight OBB vertices of $O_a$ lie inside the OBB of $O_b$, allowing
    a tolerance of $\tau_c$. The \texttt{on} relation takes precedence
    when both conditions are satisfied. &
    $\tau_c$ \\
    Under &
    \texttt{under} &
    The target center projected to XY lies inside $F_b$, the two bottoms
    are close ($|z_a^{-}-z_b^{-}|\leq\tau_u$), and the target center is
    below the reference top ($z_a<z_b^{+}$). Cases satisfying
    \texttt{on} are excluded. &
    $\tau_u$ \\
    Facing &
    \texttt{facing} &
    The target faces the reference if the angle between the target front
    direction and the vector from $O_a$ to $O_b$ is below $\tau_\theta$, with
    $d_{xy}(a,b)<\tau_d$. The front direction is derived from the target
    yaw angle. &
    $\tau_\theta,\tau_d$ \\
    \bottomrule
  \end{tabular}
\end{table*}

%% file: sections/figs/validation_loss.tex
\begin{figure}[H]
  \centering
  \includegraphics[width=\linewidth]{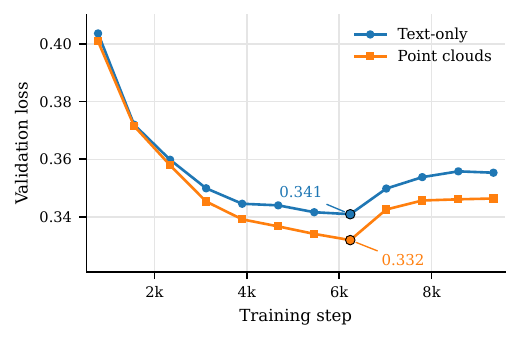}
  \caption{Validation loss curves for the text-only and point-cloud variants used in the internal comparison. Both variants reach their minimum validation loss at step 6240 (epoch 2.67), the checkpoint used for the reported test metrics; the markers annotate the corresponding validation losses.}
  \Description{Line plot showing validation loss curves for the text-only and
  point-cloud variants, with the selected minimum-loss checkpoint marked at
  step 6240 and the minimum validation losses annotated.}
  \label{fig:validation-loss}
\end{figure}

%% file: sections/tabs/artist_quality_ablation.tex
\begin{table}[!t]
  \centering
  \caption{Stratified ablation of high-quality training scenes on two held-out test subsets. The high-quality subset contains Imaginarium and artist-crafted scenes; the other contains 3D-FRONT and SpatialLM-PCG scenes. All entries are percentages.}
  \label{tab:artist-crafted-subset-ablation}
  \scriptsize
  \setlength{\tabcolsep}{2pt}
  \resizebox{\columnwidth}{!}{%
  \begin{tabular}{@{}llcccc@{}}
    \toprule
    Test subset & Method & Within. $\uparrow$ & Instruction. $\uparrow$ &
    Float. $\downarrow$ & Support. $\uparrow$ \\
    \midrule
    high-quality & Full method & \textbf{52.82} & \textbf{71.86} & \textbf{10.34} & \textbf{91.41} \\
     & w/o high-quality & 39.47 & 64.59 & 13.34 & 89.56 \\
    \addlinespace[2pt]
    3D-FRONT \& SpatialLM-PCG & Full method & \textbf{68.65} & \textbf{80.06} & 3.38 & 97.05 \\
     & w/o high-quality & 65.44 & 79.57 & \textbf{3.17} & \textbf{97.25} \\
    \bottomrule
  \end{tabular}%
  }
\end{table}